\documentclass[%
 reprint,
superscriptaddress,
nofootinbib,
 amsmath,amssymb,
 aps,
]{revtex4-2}

\usepackage{graphicx}% Include figure files
\usepackage{dcolumn}% Align table columns on decimal point
\usepackage{bm}% bold math
\usepackage{physics}
\usepackage{multirow}
\usepackage{siunitx} 
\usepackage{xcolor} % To highlight sentences when editing; remove before submission

\begin{document}

% \preprint{APS/123-QED}

% \title{Dresselhaus and Rashba spin-orbit coupling in near-surface quantum wells}
%\title{Weak antilocalization affected by the orientation of the magnetic field}
\title{Magneto-anisotropic weak antilocalization in near-surface quantum wells}

\author{S. M. Farzaneh}
\email{farzaneh@nyu.edu} 
% \affiliation{
% Physics Department, New York University, New York, NY 10003, USA}
\author{Mehdi Hatefipour}
\author{William F. Schiela}
% \affiliation{
% Physics Department, New York University, New York, NY 10003, USA}
\author{Neda Lotfizadeh}
\author{Peng Yu}

\author{Bassel Heiba Elfeky}
\author{William M. Strickland}
\affiliation{
Physics Department, New York University, New York, NY 10003, USA}
\author{Alex Matos-Abiague}
\affiliation{
Department of Physics and Astronomy, Wayne State University, Detroit, MI 48201, USA}
\author{Javad Shabani}
\email{jshabani@nyu.edu}
\affiliation{
Physics Department, New York University, New York, NY 10003, USA}

\date{November 14, 2022}

\begin{abstract}
% Spin-orbit coupling in solids describes an interaction between an electron's spin, an internal quantum-mechanical degree of freedom, and its linear momentum, an external property. 
% The strength of spin-orbit coupling, due to its relativistic nature, is typically small in solids, and is often taken into account perturbatively.
%%%%%%%%%%%%%%%%%%%%%%%%%%%%%%%%
%It has been realized, however, that semiconductors such as InAs can possess strong Rashba spin-orbit coupling where an electric field can break the structure inversion symmetry. 
%In addition, these semiconductors have a zincblende structure that results in bulk inversion asymmetry and hence Dresselhaus spin-orbit coupling. The competition between Rashba and Dresselhaus terms reduces the rotational symmetry of the effective Hamiltonian of the system. 
%The reduced symmetry of the system emerges in weak-antilocalization physics in the presence of an in-plane magnetic field. Using a semiclassical universal model, we uniquely extract values of Dresselhaus and Rashba parameters as well as the effective in-plane g-factor of the electrons.
%This is in contrast to previous works where the spin-orbit coupling is extracted from a simplified Rashba-only model which does not lead to unique solutions. 
%Understanding spin-orbit coupling parameters provides new prospects for new applications ranging from spintronics to topological quantum computing.
%%%%%%%%%%%%%%%%%%%%%%%%%%%%%
We investigate the effects of an in-plane magnetic field on the weak antilocalization signature of near-surface quantum wells lacking bulk and inversion symmetry. 
The measured magnetoconductivity exhibits a strong anisotropy with respect to the direction of the in-plane magnetic field. 
The two-fold symmetry of the observed magneto-anisotropy originates from the competition between Rashba and Dresselhaus spin-orbit couplings. 
The high sensitivity of the weak antilocalization to the spin texture produced by the combined Zeeman and spin-orbit fields results in very large anisotropy ratios, reaching 100\%. 
Using a semiclassical universal model in quantitative agreement with the experimental data, we uniquely determine the values of the Dresselhaus and Rashba parameters as well as the effective in-plane g-factor of the electrons. Understanding these parameters provides new prospects for novel applications ranging from spintronics to topological quantum computing.
\end{abstract}

\maketitle

\section{introduction}
% why spin-orbit and why InAs
Spin-orbit coupling plays a crucial role in spintronics and topological superconductivity since it provides an electronic knob to tweak the spin properties of materials. 
Moreover, quantum wells made of semiconductors, such as InAs, are considered candidates for exploring topological superconductivity due to their strong spin-orbit coupling. 
A high electron mobility along with epitaxial compatibility (along the $[001]$ direction) with Al has recently made InAs a candidate for studying nontrivial topological signatures in semiconductor-superconductor systems \cite{2021_Dartiailh, 2016_Shabani, 2017_Kjaergaard}. 
Although a strong \textit{atomic} spin-orbit coupling is a precursor to a strong \textit{effective} spin-orbit coupling in a crystalline system, the structural details of a heterostructure can greatly affect spin properties such as the effective spin-orbit strengths as well as the effective g-factor. 
In addition to the spin-orbit coupling, often an in-plane magnetic field is used to break time reversal symmetry and open up a gap in the energy spectrum of the engineered materials in order to achieve emergent properties such as non-trivial topological phases \cite{2021_Dartiailh, 2016_Shabani, 2017_Kjaergaard, 2017_Suominen, 2017_Suominena, 2019_fornieri, 2019_Ren, 2017_Pientka, 2022_Pekerten, 2021_Pakizer, 2021_Pakizera}. 
The goal of this work is to study the effect of the in-plane field and its orientation on the magnetoconductivity and the resulting weak antilocalization signature in a quasi-two-dimensional electron gas.

In an epitaxially grown InGaAs/InAs/InGaAs quantum well, we found that the magnetoconductivity can exhibit surprisingly large anisotropic ratios exceeding 100\% with respect to the orientation of the in-plane magnetic field. 
This is unexpected as there are no magnetic materials in the heterostructure. 
The lack of bulk- and structural-inversion symmetries in the heterostructure leads to the coexistence of Rashba and Dresselhaus spin-orbit coupling (SOC), which in turn, is responsible for the observed anisotropy of the magnetoconductivity. 
Modeling the experimental data with a semiclassical theory that captures the effects of both the SOC and the in-plane magnetic field, we uniquely determine the values of the in-plane g-factor and the Rashba and Dresselhaus parameters. 
This is in contrast to previous works, where the SOC was extracted from a simplified Rashba-only or Dresselhaus-only model with no unique solutions \cite{2020_Sazgari, 2017_Herling, 2012_Spirito}. 
Interestingly, the extracted in-plane g-factor is proportional to the electron density and is higher than the bulk value of InAs.

%Modeling the experimental data with a semiclassical theory, we extract the effective strength of spin-orbit coupling terms and the in-plane g-factor. 
%The spin-orbit coupling contains two terms caused by the structural inversion asymmetry (Rashba) and the bulk inversion asymmetry (Dresselhaus). 
%The inclusion of in-plane magnetic field allows us to uniquely determine Rashba and Dresselhaus parameters as well as the in-plane g-factor. 
%This is in contrast to previous works where the spin-orbit coupling is extracted from a simplified Rashba-only or Dresselhaus-only model which does not lead to unique solutions \cite{2020_Sazgari, 2017_Herling, 2012_Spirito}. 
%In our experiments, the quasi-two-dimensional system is an epitaxially grown InAs heterostructure composed of an InGaAs/InAs/InGaAs quantum well. 

% why weak antilocalization 
Weak localization and antilocalization phenomena, which are macroscopic manifestations of quantum interference \cite{1982_Poole, 1985_Kawaji, 1984_Khmelnitskii, 1986_Chakravarty} and can be thought of as quantum corrections to the magnetoconductivity, have previously been used to estimate the strength of spin-orbit coupling in semiconductors \cite{2002_Koga, 2003_Miller, 2010_Kallaher, 2011_Faniel, 2012_Spirito, 2016_Yoshizumi, 2021_Nishimura} and in particular InAs heterostructures \cite{1993_Chen, 2004_Schierholz, 2017_Herling, 2018_Wickramasinghe, 2020_Sazgari}. 
% what's missing in prior art
However, the effect on weak antilocalization due to the interplay between spin-orbit coupling and an in-plane magnetic field has not been explored in-depth before and is the focus of this work.
The spin-orbit coupling emerges as a peak in magnetoconductivity which is a result of the suppression of back scattering due to different spin states for $\vb*{k}$ and $-\vb*{k}$ momenta. 
A cryogenic temperature (few Kelvins and lower), a standard Hall bar geometry, and a relatively low magnetic field ($<$\SI{0.1}{T}) are sufficient to observe weak antilocalization in InAs heterostructures. 
Throughout the work we assume that the phase coherence length in the system is much larger than the momentum relaxation length, $\ell_\varphi \gg \ell$. 

% review of prior arts and methods 
Theoretical models of weak antilocalization are divided into two major categories depending on the relative magnetic length in the system \cite{2017_Weigele_thesis}. 
The first category of theories work in the regime of weak magnetic field where it is assumed that the magnetic length is much greater than the momentum relaxation length $\ell_{B} \gg \ell$ \cite{1980_Hikami, 1994_Iordanskii, 1996_Knap, 1995_Pikus}. 
The second category describes the regime where this assumption is relaxed or the opposite assumption is made, that is, $\ell \gg \ell_B$ \cite{1984_Kawabata, 1986_Chakravarty, 1987_Wittmann, 1994_Dyakonov, 1997_Zduniak, 2004_McPhail, 2005_Golub}. 
However some of these theories \cite{1984_Kawabata} still make an assumption about the spin-orbit length being much less than the mean free path, $\ell_\mathrm{so} \ll \ell$, such that the spins relax according to the Dyakonov-Perel mechanism. 
In our system, different length scales are comparable to each other, i.e., $\ell \sim \ell_B \sim \ell_\mathrm{so}$. Therefore, we use a universal theory that is valid in both weak and strong magnetic-field regimes as well as for any values of spin-orbit length \cite{2017_Sawada}. 
The electrons are assumed to behave semiclassically where the single-electron interactions (spin-orbit coupling and the Zeeman terms) change the phase of the electronic wavefunction via time evolution between scattering events. The magnetoconductivity is then calculated based on a real space simulation of randomly generated scattering loops in a two-dimensional plane. 
We use the same scattering loops generated in Ref. \citenum{2017_Sawada} where 147,885 predetermined loops with up to 5000 scattering events are included.

\begin{figure}[h!]
    \includegraphics[width=1.0\linewidth]{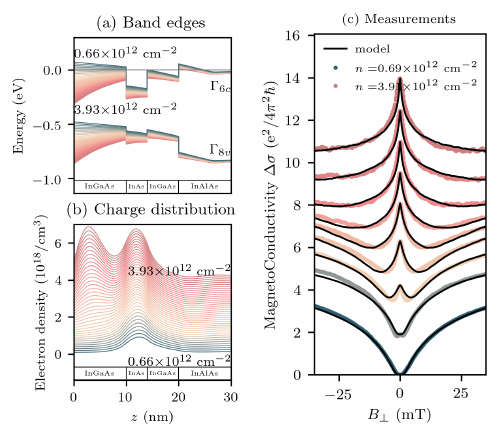}
    % \centering
    % \begingroup
    % \setlength{\tabcolsep}{0pt}
    % \renewcommand{\arraystretch}{0.2} 
    % \begin{tabular}{cc}
    %     \includegraphics[width=0.5\linewidth]{fig/density_waterfall-conduction-valence.png} & 
    %     \includegraphics[width=0.5\linewidth]{fig/js512-hb1-002-wal-waterfall-rashba-only.png} 
    % \end{tabular}
    % \endgroup 
    \caption{\label{fig:1} Calculations and experimental data: The Poisson-Schrodinger solutions of the Kane model are shown in (a) as band edges and in (b) as charge distribution. 
    Each trace corresponds to a different total density. 
    The heterostructure is made of an InAs layer sandwiched between In$_{0.81}$Ga$_{0.19}$As layers grown on In$_{0.81}$Al$_{0.19}$As. 
    The growth direction $z$ is along the $[001]$ crystallographic axis. 
    (c) Measured magnetoconductivity of the two-dimensional electron gas at different total densities. 
    The traces are shifted vertically for clarity. 
    Solid lines illustrate fits from the weak antilocalization model where only Rashba spin-orbit coupling is included.}
\end{figure}

\section{Band Simulations} 
To understand the physics of the system, the band structure of the quantum well is simulated using the extended Kane model of zinc blende semiconductors in the vicinity of the $\Gamma$ point in the momentum space. 
% We use \texttt{nextnano} \cite{2006_Trellakis} software to solve for the coupled Poisson-Schrodinger equations which self-consistently determine the electric potential and the charge distribution as well as the subbands resulting from the confinement along the growth direction $z$ ([001] crystal direction). 
% The bulk parameters of the ternary alloys, In$_{0.81}$Al$_{0.19}$As and In$_{0.81}$Ga$_{0.19}$As, are extrapolated according to Vegard's law from that of the constituting binary compounds, i.e. InAs, AlAs, and GaAs, with proper bowing parameters \cite{2001_Vurgaftman}. 
Figure \ref{fig:1} (a) and (b), respectively, show the simulated band edges ($\Gamma_6^c$, $\Gamma_8^v$) and the charge distribution along the growth direction. 
Each trace corresponds to a different total density. 
The density is assumed to be tuned by a metal-oxide structure located at $z<0$. 
The band bending at higher densities suggests that a significant Rasbha spin-orbit coupling is expected. 
Moreover, the appearance of a second hump in the charge density suggests that higher subbands become populated and therefore the Dresselhaus term proportional to $\ev{k_z^2}$ is expected to increase as the gate voltage and the density increase. 
The strength of Rashba and Dresselhaus terms can be estimated from the Poisson-Schrodinger simulations as well (SUPPLEMENTAL). 
This allows us to use the estimated spin-orbit strengths as the initial values to the fitting procedure.

\section{Magnetoconductivity Measurements}
% A gated Hall bar is fabricated on the heterostructure via photolithography and wet chemical etching. 
% The width of the Hall bar is \SI{500}{\micro m} which is assumed to be much larger than the phase and the momentum relaxation lengths as well as the Fermi wavelength so that the lateral confinement effects are negligible. 
% A \SI{30}{nm} layer of Al$_2$O$_3$ dielectric is then deposited using atomic layer deposition, followed by a \SI{70}{nm} layer of Cr/Au as the gate contact. 
The longitudinal resistance of the sample is measured as a function of a sweeping out-of-plane magnetic field $B_\perp$ (along $z$). 
The resulting magnetoconductivity is depicted in Fig. \ref{fig:1} (c) where different traces correspond to different total densities. 
At low densities we observe a signature of weak localization that is a large valley in the magnetoconductivity. As the density increases, a peak appears at $B_\perp=0$ which results from the spin-orbit coupling and signifies a weak antilocalization effect. 

\section{Fitting Procedure}
We use a semiclassical theory developed in Ref. \citenum{2017_Sawada} to capture the weak antilocalization phenomenon without making any assumptions about the magnetic length. In addition to its computational efficiency, this model can also include the effect of an in-plane magnetic field which is difficult to capture with other models. 
Moreover, no assumption is made about the spin relaxation mechanism as the time evolution of the electronic wavefunction between scattering events is calculated explicitly via the following Hamiltonian: 
\begin{equation}
\begin{split}
\label{eq:hamiltonian}
    H & = \alpha (k_y \sigma_x - k_x \sigma_y) \\
    & + \beta (k_x \sigma_x - k_y \sigma_y) 
    + \gamma (k_xk_y^2 \sigma_x - k_x^2 k_y \sigma_y) \\ 
    & + \frac{g_\parallel}{2} \mu_\mathrm{B} \vb*{B_\parallel} \cdot \vb*{\sigma}. 
\end{split}
\end{equation}
Here $\alpha$ is the strength of the Rashba spin-orbit coupling and $\beta$ and $\gamma$ denote the linearized and the cubic Dresselhaus terms. 
We have included an in-plane Zeeman term in the model to include the effect of an in-plane magnetic field $B_\parallel$ on the magnetoconductivity.
We note that the perpendicular Zeeman effect is implicitly included in the magnetoconductivity model \cite{2017_Sawada}. 
In addition to the Hamiltonian above, the magnetoconductivity depends on the phase coherence length $\ell_\varphi$ over which the energy is conserved, and, therefore, the resulting phase factor is coherently evolving for all scattering paths. 
In total, the magnetoconductivity requires five fitting parameters, i.e., $\Delta \sigma = \Delta \sigma(\ell_\varphi, \alpha, \beta, \gamma, g_\parallel)$. 
Our calculations show that there exist multiple solutions which describe the same magnetoconductivity data since the size of the peak at $B_\perp=0$ depends on the difference $\alpha - \beta$ and not the individual values of $\alpha$ and $\beta$, as others \cite{2019_Marinescu} have shown as well. 
This makes the fitting job computationally expensive which forces us to make a few simplifying assumptions. 
% the fitting recipe
Previous measurements \cite{2018_Wickramasinghe} indicate that the Rashba SOC dominates the isotropic transport properties at zero in-plane magnetic field. 
However, when the in-plane field is not zero, the competition between Rashba and Dresselhaus SOCs becomes relevant for describing the observed magneto-anisotropy induced by the in-plane field, as reported here. 
Therefore, we first consider the isotropic effects at zero in-plane field ($B_\parallel =0$), where the Rashba term is dominant and the Dresselhaus parameters can be set to zero (i.e., $\beta = 0$, $\gamma = 0$).
We then fit $\Delta\sigma$ using only two parameters, $\ell_\varphi$ and $\alpha$. 
The resulting fits are depicted in Fig. \ref{fig:1} (c). 
These fits are able to capture the magnetoconductivity data at zero in-plane field, even though only the Rashba term is included in the model. 
The fits are in accordance with what has been shown in previous transport measurements as well \cite{2018_Wickramasinghe}. 
To verify the validity of the weak antilocalization model at zero in-plane field, the values of $\alpha$ are compared against those extracted from a commonly used model in the literature by Knap et al. \cite{1996_Knap} (SUPPLEMENTAL).
Nevertheless, it should be noted that the model by Knap et al. is based on the Dyakonov-Perel mechanism requiring that $\ell_\mathrm{so} < \ell$ which in our system corresponds to densities lower than $3.5\mathrm{e}12$ cm$^{-2}$.
In the presence of an in-plane magnetic field, however, the Rashba-only model fails to describe the experimental data. 

\section{In-plane Magnetic field}
\begin{figure*}[ht]
    \includegraphics[width=1.0\linewidth]{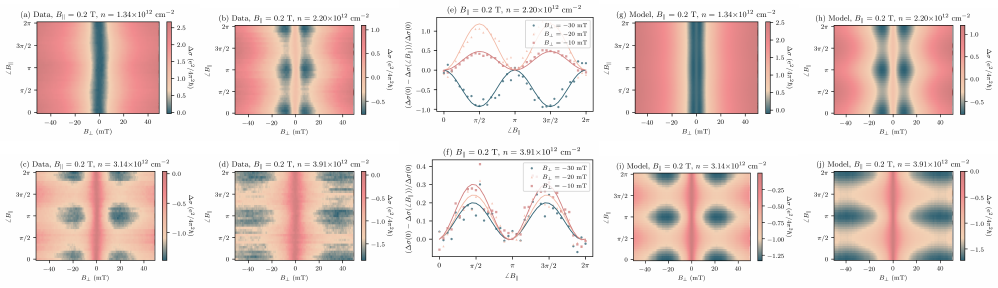}
    % \centering
    % \begingroup
    % \setlength{\tabcolsep}{0pt}
    % \renewcommand{\arraystretch}{0.2} 
    % \begin{tabular}{ccccc}
    %     \includegraphics[width=0.195\linewidth]{fig/js512-hb1-002-inplane-angle-1-200mT-a.png} & 
    %     \includegraphics[width=0.195\linewidth]{fig/js512-hb1-002-inplane-angle-3-200mT-b.png} & 
    %     \includegraphics[width=0.195\linewidth]{fig/js512-hb1-fit-inplane-field-cut-3-200mT-e.png} & 
    %     \includegraphics[width=0.195\linewidth]{fig/js512-hb1-002-model-angle-1-g.png} & 
    %     \includegraphics[width=0.195\linewidth]{fig/js512-hb1-002-model-angle-3-h.png} 
    %     \\ 
    %     \includegraphics[width=0.195\linewidth]{fig/js512-hb1-002-inplane-angle-5-200mT-c.png} & \includegraphics[width=0.195\linewidth]{fig/js512-hb1-002-inplane-angle-7-200mT-d.png} &
    %     \includegraphics[width=0.195\linewidth]{fig/js512-hb1-fit-inplane-field-cut-7-200mT-f.png} & 
    %     \includegraphics[width=0.195\linewidth]{fig/js512-hb1-002-model-angle-5-i.png} & \includegraphics[width=0.195\linewidth]{fig/js512-hb1-002-model-angle-7-j.png} \\
    % \end{tabular}
    % \endgroup
        \caption{\label{fig:2} (a)--(d) Magnetoconductivity, $\Delta \sigma$, versus different orientations of the in-plane magnetic field with a fixed magnitude of $B_\parallel = $ \SI{0.2}{T} for different densities. The data is manually shifted so that $\Delta \sigma = 0$ at $B_\perp = 0$. The angle $\angle B_\parallel$ is defined with respect to the [110] crystal axis. Panels (e) and (f) show vertical cuts which are normalized and show a $\cos(2\angle B_\parallel)$ dependence. (g)--(j) Magnetoconductivity reproduced using the semiclassical theory of weak antilocalization.} 
\end{figure*}

In the presence of an in-plane magnetic field $B_\parallel$, we observe that the magnetoconductivity depends on the orientation of the field with respect to the crystal axes, $\angle B_\parallel$. 
Figures \ref{fig:2}(a)-(d) illustrate the measured magnetoconductivity $\Delta \sigma (\angle B_\parallel)$ as a function of the orientation of the in-plane field for four different densities. 
The orientation $\angle B_\parallel = 0$ corresponds to the [110] crystal axis which is rotated by $45$ degrees with respect to the $x$ axis in the Hamiltonian of Eq. \ref{eq:hamiltonian}. 
The magneto-anisotropic effects manifest in the lack of rotational symmetry with respect to $\angle B_\parallel$, which is particularly apparent in the higher density cases. 
To better quantify the strength of the anisotropy, we introduce the magneto-anisotropy ratio, $\big(\Delta\sigma(0)-\Delta\sigma(\angle B_\parallel))/\Delta\sigma(0)$. 
The magneto-anisotropy ratio as a function of the in-plane field direction is shown in Figs.~\ref{fig:2}(e) and (f). 
Symbols represent the experimental data and solid lines correspond to fittings using a phenomenological model capturing the interrelation between in-plane field and the Rashba and Dresselhaus SOCs (SUPPLEMENTAL). 
When $\angle B_\parallel = \pi/2$, the corresponding magneto-anisotropy ratio measures the relative change in the magnetoconductivity when the in-plane field orientation is rotated from the $[110]$ to the $[\bar{1}10]$ crystallographic direction. 
These are precisely the symmetry axes of the two-fold symmetric spin-orbit field resulting from the coexistence of Rashba and Dresselhaus SOCs, and correspond to directions along which the total spin-orbit field strength reaches its extreme (maximum or minimum) values. 
Therefore the temperature-independent dephasing rate becomes anisotropic \cite{1999_Malshukov} and 
the destructive interference between electron paths in opposite directions along a loop strongly depends on whether the in-plane field is aligned with the direction of maximum or minimum spin-orbit field strength. 
As a result, the weak antilocalization effect becomes highly sensitive to the in-plane field direction, leading to the large magneto-anisotropy ratios ($~20\%-100\%$) shown in Figs.~\ref{fig:2}(e) and (f) for $\angle B_\parallel = \pi/2,3\pi/2$. 
Even though the considered heterostructure does not contain magnetic materials, the measured anisotropic ratios are comparable and even larger than those detected in magneto-anisotropic phenomena involving magnetic elements and/ or in the presence of strain \cite{2015_Hupfauer, 2014_Jungwirth, 2003_Loehr}.

% \onecolumngrid
% \twocolumngrid
The Rashba spin-orbit coupling alone, due to its rotational symmetry, cannot explain the experimental data in the presence of $B_\parallel$. 
The coexistence of Rashba and Dresselhaus SOCs reduces the spin-orbit field symmetry to $C_{2v}$, which is crucial for the emergence of the magneto-anisotropy. 
Therefore, by including both Rashba ($\alpha\neq 0$) and Dresselhaus ($\beta\neq 0$ and $\gamma\neq 0$) terms, we are able to capture the effect of the in-plane field orientation. 
Using a semiclassical theory of weak antilocalization, we computed the magnetoconductivity shown in Figs. \ref{fig:2}(g)-(i) for different densities. 
The theoretical model is in good agreement with the experimental data shown in  Figs. \ref{fig:2}(a)-(d) and is further used for simultaneously fitting the spin-orbit coupling strengths $\alpha$, $\beta$, $\gamma$, and the in-plane g-factor, $g_\parallel$.
The starting values of $\alpha$ are obtained from the Rashba-only fits shown in Fig. \ref{fig:1}c. 
The starting values of $\beta$ and $\gamma$ are obtained from the solutions of Poisson-Schrodinger equations (SUPPLEMENTAL). The values of $\alpha$, $\beta$, and $\gamma$ are then tweaked in the presence of a non-zero $g_\parallel$ so that the resulting magnetoconductivity captures the experimental data in the presence of $B_\parallel$. 
This allows for a unique and consistent determination of these important parameters, which are relevant for spintronics \cite{2004_Zutic, 2007_Fabian} and topological superconductivity \cite{2021_Dartiailh, 2016_Shabani, 2017_Kjaergaard, 2017_Suominen, 2017_Suominena, 2019_fornieri, 2019_Ren, 2017_Pientka, 2022_Pekerten, 2021_Pakizer, 2021_Pakizera}. 

\section{Extracting g-factor}
\begin{figure}[ht]
    \includegraphics[width=1.0\linewidth]{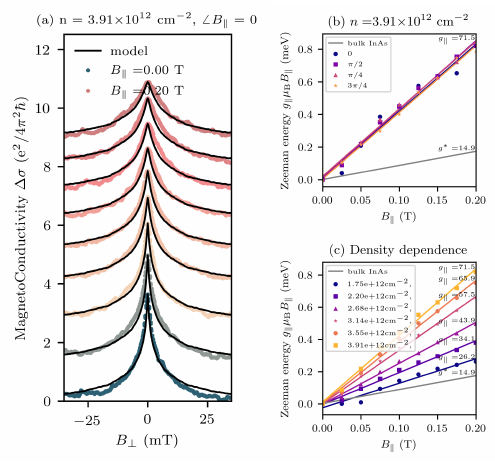}
    % \centering
    % \begingroup
    % \setlength{\tabcolsep}{0pt}
    % \renewcommand{\arraystretch}{0.2} 
    % \begin{tabular}{cc}
    %     \multirow{2}{*}{\includegraphics[width=0.54\linewidth]{fig/js512-hb1-002-wal-waterfall-inplane-7-0.png}} \\
    %     & 
    %     \includegraphics[width=0.46\linewidth]{fig/js512-hb1-002-zeeman-7-beta.png} \\
    %     & 
    %     \includegraphics[width=0.46\linewidth]{fig/js512-hb1-002-gfactor-density.png}
    % \end{tabular}
    % \endgroup
    \caption{\label{fig:3} 
        (a) Magnetoconductivity of the two-dimensional electron gas in the presence of an in-plane magnetic field $B_\parallel$ with different magnitudes with a fixed density and a fixed $\angle B_\parallel$. The in-plane field diminishes the peak in the magnetoconductivity. The traces are shifted vertically for clarity. Solid lines illustrate fits from the model. 
        (b) Zeeman energy as a function of $B_\parallel$ for different orientations $\angle B_\parallel$ extracted from fitting different $\Delta \sigma$ traces. The Dresselhaus parameter, $\beta=$ \SI{19}{meV\AA}, is the same for different orientations. The solid line is the linear fit and the slope determines the g-factor.
        (c) Zeeman energy vs. $B_\parallel$ for different densities.
        }
\end{figure}

\begin{figure}[ht]
    \includegraphics[width=0.99\linewidth]{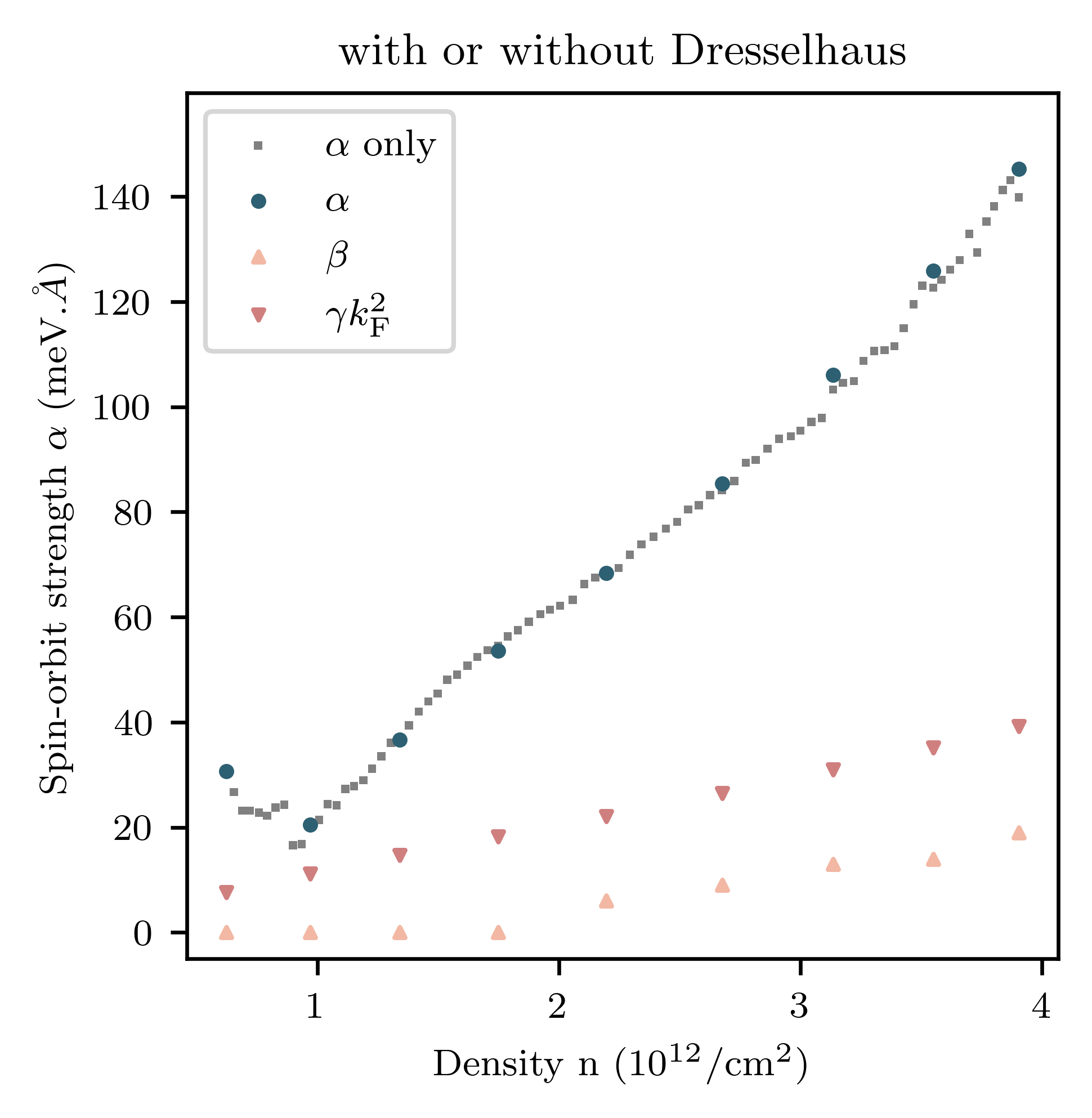} 
    \caption{\label{fig:4} 
        Comparing strengths of spin-orbit coupling terms with or without Dresselhaus terms. Square marker denotes extracted values when there is only Rashba term included in the Hamiltonian. Other markers denote the fit values in the presence of both Rashba and Dresselhaus terms.}
\end{figure}

To extract the in-plane electron g-factor, we measure the magnetoconductivity for different magnitudes of $B_\parallel$ as shown in Fig. \ref{fig:3} (a), where different traces correspond to different values of $B_\parallel$. 
As the magnitude of $B_\parallel$ increases the peak of weak antilocalization gets diminished. 
The reason is that the in-plane magnetic field introduces a coherent rotation on the spin of all electrons irrespective of their momentum and therefore interferes with the phase induced by the spin-orbit coupling terms. Consequently the spins tend to align along the in-plane field which diminishes the weak antilocalization signature. 
The solid lines demonstrate the fits from the model. 
Each fit corresponds to a value of Zeeman energy $g_\parallel \mu_\mathrm{B} B_\parallel$ which is plotted as a function of $B_\parallel$ in Fig. \ref{fig:3} (b).
Not surprisingly, the Zeeman energy is linear in $B_\parallel$ and $g_\parallel$ is extracted from the slope of a linear fit. 
Different colors in Fig. \ref{fig:3} (b) correspond to different $\angle B_\parallel$. 
The main point is that when $\beta=0$ the slope of the linear fits are not the same which leads to different values of $g_\parallel$ for different directions. However, by including a non-zero $\beta\not=0$ in the model and simultaneously fitting $\beta$ and $g_\parallel$ we are able to obtain linear fits with the same slope for four different orientations, i.e., $\angle B_\parallel = 0, \pi/4, \pi/2, 3\pi/4$ which results in a $g_\parallel = 71.5$ at the density $n=$ \SI{3.91e12}{cm^{-2}}. 
In this case, $\beta=$ \SI{19}{meV\AA} uniquely captures the anisotropic behavior of magnetoconductivity vs $\angle B_\parallel$. 
This value is much smaller than the Rashba strength $\alpha=$ \SI{150}{meV\AA}. 
The procedure is repeated for different densities and the resulting Zeeman energies and $g_\parallel$ are shown in Fig. \ref{fig:3} (c). 
Once the values of $\beta$ and $g_\parallel$ are obtained, the orientation-dependent magnetoconductivity data can be captured by the model for the whole range of $\angle B_\parallel$ as shown in Fig. \ref{fig:2}. 
The values of spin-orbit coupling strength for both cases with and without Dresselhaus terms are plotted in Fig. \ref{fig:4}. 

\section{Length scales}
To verify the validity of our assumptions, we list the relevant energy, length, and field scales of different parameters in Table \ref{tab:scale}. 
\begin{table}[ht]
\caption{Relevant field, length, and energy scales of the system at $T = $ \SI{40}{mK}. Here we use the bulk g-factor of InAs, that is $g^*=14.9$ and $m^*=0.04m_e$ as the effective electron mass \cite{2020_Yuan}. Wherever the parameters take a range of values we report the value at the fixed density of $n=$ \SI{2e12}{cm^{-2}}. The length scale $\ell$ and the magnetic field $B$ are related via $\ell = \sqrt{\hbar/eB}$.}
\label{tab:scale} 
\centering
\begin{tabular}{lrrr}
\hline
        Quantity & Energy (meV) & Length (nm) & Field (mT) \\
\hline
     $g^*\mu_\mathrm{B}B$ & $0.862$ & $25.7$ & 1000 \\
     $k_\mathrm{B}T$ & 0.003 & & \\
     $\hbar^2k_\mathrm{F}^2/2m^*$ & 120 & $2\pi/k_\mathrm{F}=17.7$ & 2100 \\
     $\hbar^2\ev{k_z^2}/2m^*$ & 38.1 & $2\pi/\sqrt{\ev{k_z^2}}=31.4$ & 667 \\
     $\ell$ & $\hbar/\tau = 6.38$ &  106 & 58.7 \\
     $\ell_\varphi$ & $\hbar/\tau_\varphi = 0.032$ & 20900 & 0.002 \\
     $\alpha$ & $\alpha k_\mathrm{F} = 2.25$ & $\hbar^2/2m^*\alpha=150$ & 29.2 \\
     $\beta$ & $\beta k_\mathrm{F} = 0.303$ & $\hbar^2/2m^*\beta=1120$ & 0.528 \\
     $\gamma$ & $\gamma k_\mathrm{F}^3 = 1.215$ & $\hbar^2/2m^*\gamma k_\mathrm{F}^2 = 278$ & 8.50 \\
     \hline
     \multicolumn{4}{l}{$\alpha$: Rashba, $\beta$, $\gamma$: linearized and cubic Dresselhaus}\\
     \multicolumn{4}{l}{$\ell$, $\tau$: momentum relaxation length and time}\\
     \multicolumn{4}{l}{$\ell_\varphi$,    $\tau_\varphi$: phase coherence length and time}
\end{tabular}
\end{table}
One of the implicit assumptions of the weak antilocalization model is that the Fermi sea is isotropic which is valid as the Fermi energy is two orders of magnitude greater than the spin-orbit coupling energies, that is $\epsilon_\mathrm{F} \gg \alpha k_\mathrm{F}, \beta k_\mathrm{F}, \gamma k_\mathrm{F}^3$. 
Another key assumption is that the temperature is low enough so that the phase coherence length $\ell_\varphi = $ \SI{20.9}{\micro m} is much greater than any other length scales especially the momentum relaxation length $\ell = $ \SI{106}{nm} and the spin-orbit length $\ell_\mathrm{so} = \hbar^2/2m^*\alpha = $ \SI{150}{nm}. 
This is a precondition to observing any quantum interference effect. 
We note that since $\ell > \ell_B$, theories based on the assumption of weak magnetic field cannot describe the magnetoconductivity data accurately. 
Moreover, since the mean free path $\ell$ is comparable to the spin-orbit length $\ell_\mathrm{so}$, the relaxation of spin cannot be accurately described by the Dyakonov-Perel mechanism which requires $\ell \ll \ell_\mathrm{so}$. 

\section{Conclusions}\label{conclusion}
In the present work, we investigated the effects of an in-plane magnetic field on the magnetoconductivity of a near-surface semiconductor quantum well in the weak antilocalization regime. 
The experimental results reported here reveal the high sensitivity of the weak antilocalization to the orientation of the in-plane magnetic field, yielding a strongly anisotropic magnetoconductivity. 
The magneto-anisotropy exhibits a two-fold symmetry attributed to the interplay between Rashba and Dresselhaus SOCs. 
We showed that our  experimental magnetoconductivity data can be properly modeled by a semiclassical theory which explicitly includes the magnetic field and the SOC effects, and allows for simultaneously fitting the in-plane g-factor and the Rashba and Dresselahus SOC strengths to the data. 
Unlike previous studies ignoring the in-plane anisotropic effects \cite{2004_Minkov}, our systematic procedure provides a consistent way of uniquely determining the SOC parameters. 

The extracted values of the SOC strengths confirm the dominance of the Rashba contribution, as has been shown in similar structures \cite{2000_Grundler}. 
The smallest contribution corresponds to the linearized Dresselhaus term, which is found to be almost an order of magnitude smaller than the Rashba SOC strength. 
We found that the anisotropy of the Zeeman interaction, which is possible in non-symmetric quantum wells \cite{1996_Oestreich}, is fairly small compared to the observed magnetoconductivity anisotropy, which is mainly determined by the coexistence of Rashba and Dresselhaus SOCs. 
The extracted values of the electron's in-plane g-factor are surprisingly higher than that of bulk InAs by a factor of 2 to 5. This can be explained by the steep band bending at the surface caused by the high electron density. 
A higher electron density corresponds to a lower conduction band because of the band bending which in turn results in a lowered effective band gap which entails a larger g-factor according to the Roth formula for semiconductors.
Further investigation is required to reach a conclusive description of the g-factor behavior vs density.

% \textcolor{purple}{PLEASE, REVISE THIS COMMENT AND MOVE IT TO SECTION IV? 
% In this work we introduce a method to extract the strength of Dresselhaus spin-orbit coupling as well as the electron's in-plane g-factor, $g_\parallel$, from the magnetoconductivity in the weak antilocalization regime. 
% Previous work had modeled the disappearance of the antilocalization signature by an effective phase coherence length dependent on the in-plane magnetic field \cite{2004_Minkov}. 
% Although this model fits the experimental data, its physical origin is unclear as one would assume that the phase coherence is dominated by thermal effects and not the in-plane magnetic field.}

\section*{Methods}\label{method}
\subsection*{Simulations}
We use \texttt{nextnano} \cite{2006_Trellakis} software to solve for the coupled Poisson-Schrodinger equations which self-consistently determine the electric potential and the charge distribution as well as the subbands resulting from the confinement along the growth direction $z$ ([001] crystal direction). 
The bulk parameters of the ternary alloys, In$_{0.81}$Al$_{0.19}$As and In$_{0.81}$Ga$_{0.19}$As, are extrapolated according to Vegard's law from that of the constituting binary compounds, i.e. InAs, AlAs, and GaAs, with proper bowing parameters \cite{2001_Vurgaftman}. 

\subsection*{Hall bar}
A gated L-shaped Hall bar is fabricated on the heterostructure via photolithography and wet chemical etching. 
The width of the Hall bar is \SI{150}{\micro m} which is assumed to be much larger than the phase and the momentum relaxation lengths as well as the Fermi wavelength so that the lateral confinement effects are negligible. 
A \SI{30}{nm} layer of Al$_2$O$_3$ dielectric is then deposited using atomic layer deposition, followed by a \SI{70}{nm} layer of Cr/Au as the gate contact. 

\begin{figure}[ht]
    \includegraphics[width=0.5\linewidth]{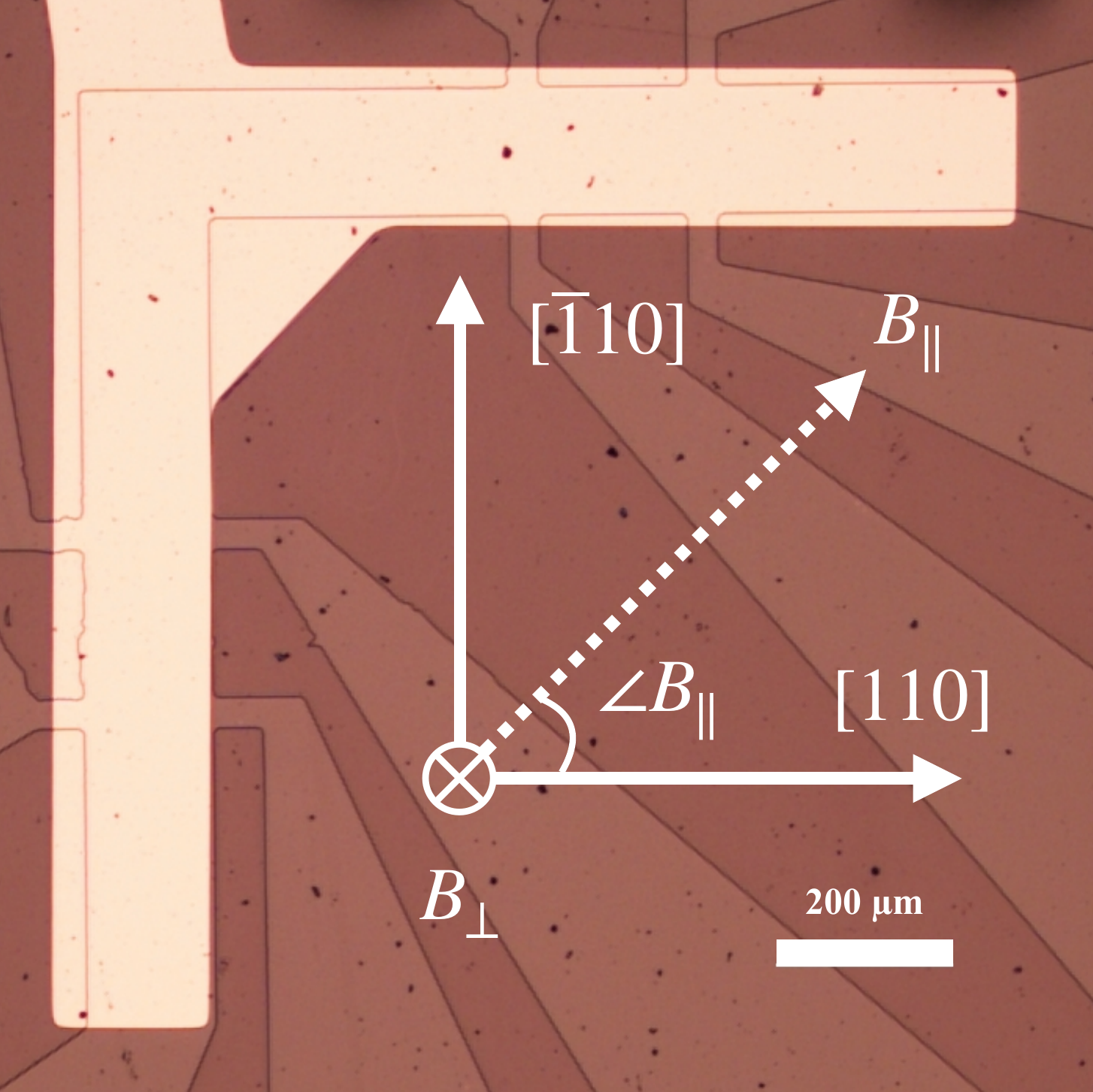} 
    \caption{\label{fig:5} 
        The optical image of the fabricated L-shaped Hall bar. The crystal axes along with the orientation of the external magnetic fields are denoted as well.}
\end{figure}

\begin{acknowledgments}
This work is supported in part by DARPA Topological Excitations in Electronics (TEE) program under grant No. DP18AP900007, the ONR grant N00014-21-1-2450, and the DOE grant DE-SC0022245.
William M. Strickland acknowledges funding from the ARO/LP  S QuaCGR fellowship reference W911NF2110303. 
William F. Schiela acknowledges funding from the U.S. Department of Defense NDSEG Fellowship.
\end{acknowledgments}

\bibliography{_manuscript_wal}

\begin{thebibliography}{54}%
\makeatletter
\providecommand \@ifxundefined [1]{%
 \@ifx{#1\undefined}
}%
\providecommand \@ifnum [1]{%
 \ifnum #1\expandafter \@firstoftwo
 \else \expandafter \@secondoftwo
 \fi
}%
\providecommand \@ifx [1]{%
 \ifx #1\expandafter \@firstoftwo
 \else \expandafter \@secondoftwo
 \fi
}%
\providecommand \natexlab [1]{#1}%
\providecommand \enquote  [1]{``#1''}%
\providecommand \bibnamefont  [1]{#1}%
\providecommand \bibfnamefont [1]{#1}%
\providecommand \citenamefont [1]{#1}%
\providecommand \href@noop [0]{\@secondoftwo}%
\providecommand \href [0]{\begingroup \@sanitize@url \@href}%
\providecommand \@href[1]{\@@startlink{#1}\@@href}%
\providecommand \@@href[1]{\endgroup#1\@@endlink}%
\providecommand \@sanitize@url [0]{\catcode `\\12\catcode `\$12\catcode
  `\&12\catcode `\#12\catcode `\^12\catcode `\_12\catcode `\%12\relax}%
\providecommand \@@startlink[1]{}%
\providecommand \@@endlink[0]{}%
\providecommand \url  [0]{\begingroup\@sanitize@url \@url }%
\providecommand \@url [1]{\endgroup\@href {#1}{\urlprefix }}%
\providecommand \urlprefix  [0]{URL }%
\providecommand \Eprint [0]{\href }%
\providecommand \doibase [0]{https://doi.org/}%
\providecommand \selectlanguage [0]{\@gobble}%
\providecommand \bibinfo  [0]{\@secondoftwo}%
\providecommand \bibfield  [0]{\@secondoftwo}%
\providecommand \translation [1]{[#1]}%
\providecommand \BibitemOpen [0]{}%
\providecommand \bibitemStop [0]{}%
\providecommand \bibitemNoStop [0]{.\EOS\space}%
\providecommand \EOS [0]{\spacefactor3000\relax}%
\providecommand \BibitemShut  [1]{\csname bibitem#1\endcsname}%
\let\auto@bib@innerbib\@empty
%</preamble>
\bibitem [{\citenamefont {Dartiailh}\ \emph {et~al.}(2021)\citenamefont
  {Dartiailh}, \citenamefont {Mayer}, \citenamefont {Yuan}, \citenamefont
  {Wickramasinghe}, \citenamefont {Matos-Abiague}, \citenamefont
  {{\v{Z}}uti{\'{c}}},\ and\ \citenamefont {Shabani}}]{2021_Dartiailh}%
  \BibitemOpen
  \bibfield  {author} {\bibinfo {author} {\bibfnamefont {M.~C.}\ \bibnamefont
  {Dartiailh}}, \bibinfo {author} {\bibfnamefont {W.}~\bibnamefont {Mayer}},
  \bibinfo {author} {\bibfnamefont {J.}~\bibnamefont {Yuan}}, \bibinfo {author}
  {\bibfnamefont {K.~S.}\ \bibnamefont {Wickramasinghe}}, \bibinfo {author}
  {\bibfnamefont {A.}~\bibnamefont {Matos-Abiague}}, \bibinfo {author}
  {\bibfnamefont {I.}~\bibnamefont {{\v{Z}}uti{\'{c}}}},\ and\ \bibinfo
  {author} {\bibfnamefont {J.}~\bibnamefont {Shabani}},\ }\bibfield  {title}
  {\bibinfo {title} {Phase signature of topological transition in josephson
  junctions},\ }\href {https://doi.org/10.1103/physrevlett.126.036802}
  {\bibfield  {journal} {\bibinfo  {journal} {Physical Review Letters}\
  }\textbf {\bibinfo {volume} {126}},\ \bibinfo {pages} {036802} (\bibinfo
  {year} {2021})}\BibitemShut {NoStop}%
\bibitem [{\citenamefont {Shabani}\ \emph {et~al.}(2016)\citenamefont
  {Shabani}, \citenamefont {Kjaergaard}, \citenamefont {Suominen},
  \citenamefont {Kim}, \citenamefont {Nichele}, \citenamefont {Pakrouski},
  \citenamefont {Stankevic}, \citenamefont {Lutchyn}, \citenamefont
  {Krogstrup}, \citenamefont {Feidenhans{\textquotesingle}l}, \citenamefont
  {Kraemer}, \citenamefont {Nayak}, \citenamefont {Troyer}, \citenamefont
  {Marcus},\ and\ \citenamefont {Palmstr{\o}m}}]{2016_Shabani}%
  \BibitemOpen
  \bibfield  {author} {\bibinfo {author} {\bibfnamefont {J.}~\bibnamefont
  {Shabani}}, \bibinfo {author} {\bibfnamefont {M.}~\bibnamefont {Kjaergaard}},
  \bibinfo {author} {\bibfnamefont {H.~J.}\ \bibnamefont {Suominen}}, \bibinfo
  {author} {\bibfnamefont {Y.}~\bibnamefont {Kim}}, \bibinfo {author}
  {\bibfnamefont {F.}~\bibnamefont {Nichele}}, \bibinfo {author} {\bibfnamefont
  {K.}~\bibnamefont {Pakrouski}}, \bibinfo {author} {\bibfnamefont
  {T.}~\bibnamefont {Stankevic}}, \bibinfo {author} {\bibfnamefont {R.~M.}\
  \bibnamefont {Lutchyn}}, \bibinfo {author} {\bibfnamefont {P.}~\bibnamefont
  {Krogstrup}}, \bibinfo {author} {\bibfnamefont {R.}~\bibnamefont
  {Feidenhans{\textquotesingle}l}}, \bibinfo {author} {\bibfnamefont
  {S.}~\bibnamefont {Kraemer}}, \bibinfo {author} {\bibfnamefont
  {C.}~\bibnamefont {Nayak}}, \bibinfo {author} {\bibfnamefont
  {M.}~\bibnamefont {Troyer}}, \bibinfo {author} {\bibfnamefont {C.~M.}\
  \bibnamefont {Marcus}},\ and\ \bibinfo {author} {\bibfnamefont {C.~J.}\
  \bibnamefont {Palmstr{\o}m}},\ }\bibfield  {title} {\bibinfo {title}
  {Two-dimensional epitaxial superconductor-semiconductor heterostructures: A
  platform for topological superconducting networks},\ }\href
  {https://doi.org/10.1103/physrevb.93.155402} {\bibfield  {journal} {\bibinfo
  {journal} {Physical Review B}\ }\textbf {\bibinfo {volume} {93}},\ \bibinfo
  {pages} {155402} (\bibinfo {year} {2016})}\BibitemShut {NoStop}%
\bibitem [{\citenamefont {Kjaergaard}\ \emph {et~al.}(2017)\citenamefont
  {Kjaergaard}, \citenamefont {Suominen}, \citenamefont {Nowak}, \citenamefont
  {Akhmerov}, \citenamefont {Shabani}, \citenamefont {Palmstr{\o}m},
  \citenamefont {Nichele},\ and\ \citenamefont {Marcus}}]{2017_Kjaergaard}%
  \BibitemOpen
  \bibfield  {author} {\bibinfo {author} {\bibfnamefont {M.}~\bibnamefont
  {Kjaergaard}}, \bibinfo {author} {\bibfnamefont {H.}~\bibnamefont
  {Suominen}}, \bibinfo {author} {\bibfnamefont {M.}~\bibnamefont {Nowak}},
  \bibinfo {author} {\bibfnamefont {A.}~\bibnamefont {Akhmerov}}, \bibinfo
  {author} {\bibfnamefont {J.}~\bibnamefont {Shabani}}, \bibinfo {author}
  {\bibfnamefont {C.}~\bibnamefont {Palmstr{\o}m}}, \bibinfo {author}
  {\bibfnamefont {F.}~\bibnamefont {Nichele}},\ and\ \bibinfo {author}
  {\bibfnamefont {C.}~\bibnamefont {Marcus}},\ }\bibfield  {title} {\bibinfo
  {title} {Transparent semiconductor-superconductor interface and induced gap
  in an epitaxial heterostructure josephson junction},\ }\href
  {https://doi.org/10.1103/physrevapplied.7.034029} {\bibfield  {journal}
  {\bibinfo  {journal} {Physical Review Applied}\ }\textbf {\bibinfo {volume}
  {7}},\ \bibinfo {pages} {034029} (\bibinfo {year} {2017})}\BibitemShut
  {NoStop}%
\bibitem [{\citenamefont {Suominen}\ \emph
  {et~al.}(2017{\natexlab{a}})\citenamefont {Suominen}, \citenamefont {Danon},
  \citenamefont {Kjaergaard}, \citenamefont {Flensberg}, \citenamefont
  {Shabani}, \citenamefont {Palmstr{\o}m}, \citenamefont {Nichele},\ and\
  \citenamefont {Marcus}}]{2017_Suominen}%
  \BibitemOpen
  \bibfield  {author} {\bibinfo {author} {\bibfnamefont {H.~J.}\ \bibnamefont
  {Suominen}}, \bibinfo {author} {\bibfnamefont {J.}~\bibnamefont {Danon}},
  \bibinfo {author} {\bibfnamefont {M.}~\bibnamefont {Kjaergaard}}, \bibinfo
  {author} {\bibfnamefont {K.}~\bibnamefont {Flensberg}}, \bibinfo {author}
  {\bibfnamefont {J.}~\bibnamefont {Shabani}}, \bibinfo {author} {\bibfnamefont
  {C.~J.}\ \bibnamefont {Palmstr{\o}m}}, \bibinfo {author} {\bibfnamefont
  {F.}~\bibnamefont {Nichele}},\ and\ \bibinfo {author} {\bibfnamefont {C.~M.}\
  \bibnamefont {Marcus}},\ }\bibfield  {title} {\bibinfo {title} {Anomalous
  fraunhofer interference in epitaxial superconductor-semiconductor josephson
  junctions},\ }\href {https://doi.org/10.1103/physrevb.95.035307} {\bibfield
  {journal} {\bibinfo  {journal} {Physical Review B}\ }\textbf {\bibinfo
  {volume} {95}},\ \bibinfo {pages} {035307} (\bibinfo {year}
  {2017}{\natexlab{a}})}\BibitemShut {NoStop}%
\bibitem [{\citenamefont {Suominen}\ \emph
  {et~al.}(2017{\natexlab{b}})\citenamefont {Suominen}, \citenamefont
  {Kjaergaard}, \citenamefont {Hamilton}, \citenamefont {Shabani},
  \citenamefont {Palmstr{\o}m}, \citenamefont {Marcus},\ and\ \citenamefont
  {Nichele}}]{2017_Suominena}%
  \BibitemOpen
  \bibfield  {author} {\bibinfo {author} {\bibfnamefont {H.}~\bibnamefont
  {Suominen}}, \bibinfo {author} {\bibfnamefont {M.}~\bibnamefont
  {Kjaergaard}}, \bibinfo {author} {\bibfnamefont {A.}~\bibnamefont
  {Hamilton}}, \bibinfo {author} {\bibfnamefont {J.}~\bibnamefont {Shabani}},
  \bibinfo {author} {\bibfnamefont {C.}~\bibnamefont {Palmstr{\o}m}}, \bibinfo
  {author} {\bibfnamefont {C.}~\bibnamefont {Marcus}},\ and\ \bibinfo {author}
  {\bibfnamefont {F.}~\bibnamefont {Nichele}},\ }\bibfield  {title} {\bibinfo
  {title} {Zero-energy modes from coalescing andreev states in a
  two-dimensional semiconductor-superconductor hybrid platform},\ }\href
  {https://doi.org/10.1103/physrevlett.119.176805} {\bibfield  {journal}
  {\bibinfo  {journal} {Physical Review Letters}\ }\textbf {\bibinfo {volume}
  {119}},\ \bibinfo {pages} {176805} (\bibinfo {year}
  {2017}{\natexlab{b}})}\BibitemShut {NoStop}%
\bibitem [{\citenamefont {Fornieri}\ \emph {et~al.}(2019)\citenamefont
  {Fornieri}, \citenamefont {Whiticar}, \citenamefont {Setiawan}, \citenamefont
  {Portol{\'{e}}s}, \citenamefont {Drachmann}, \citenamefont {Keselman},
  \citenamefont {Gronin}, \citenamefont {Thomas}, \citenamefont {Wang},
  \citenamefont {Kallaher}, \citenamefont {Gardner}, \citenamefont {Berg},
  \citenamefont {Manfra}, \citenamefont {Stern}, \citenamefont {Marcus},\ and\
  \citenamefont {Nichele}}]{2019_fornieri}%
  \BibitemOpen
  \bibfield  {author} {\bibinfo {author} {\bibfnamefont {A.}~\bibnamefont
  {Fornieri}}, \bibinfo {author} {\bibfnamefont {A.~M.}\ \bibnamefont
  {Whiticar}}, \bibinfo {author} {\bibfnamefont {F.}~\bibnamefont {Setiawan}},
  \bibinfo {author} {\bibfnamefont {E.}~\bibnamefont {Portol{\'{e}}s}},
  \bibinfo {author} {\bibfnamefont {A.~C.~C.}\ \bibnamefont {Drachmann}},
  \bibinfo {author} {\bibfnamefont {A.}~\bibnamefont {Keselman}}, \bibinfo
  {author} {\bibfnamefont {S.}~\bibnamefont {Gronin}}, \bibinfo {author}
  {\bibfnamefont {C.}~\bibnamefont {Thomas}}, \bibinfo {author} {\bibfnamefont
  {T.}~\bibnamefont {Wang}}, \bibinfo {author} {\bibfnamefont {R.}~\bibnamefont
  {Kallaher}}, \bibinfo {author} {\bibfnamefont {G.~C.}\ \bibnamefont
  {Gardner}}, \bibinfo {author} {\bibfnamefont {E.}~\bibnamefont {Berg}},
  \bibinfo {author} {\bibfnamefont {M.~J.}\ \bibnamefont {Manfra}}, \bibinfo
  {author} {\bibfnamefont {A.}~\bibnamefont {Stern}}, \bibinfo {author}
  {\bibfnamefont {C.~M.}\ \bibnamefont {Marcus}},\ and\ \bibinfo {author}
  {\bibfnamefont {F.}~\bibnamefont {Nichele}},\ }\bibfield  {title} {\bibinfo
  {title} {Evidence of topological superconductivity in planar josephson
  junctions},\ }\href {https://doi.org/10.1038/s41586-019-1068-8} {\bibfield
  {journal} {\bibinfo  {journal} {Nature}\ }\textbf {\bibinfo {volume} {569}},\
  \bibinfo {pages} {89} (\bibinfo {year} {2019})}\BibitemShut {NoStop}%
\bibitem [{\citenamefont {Ren}\ \emph {et~al.}(2019)\citenamefont {Ren},
  \citenamefont {Pientka}, \citenamefont {Hart}, \citenamefont {Pierce},
  \citenamefont {Kosowsky}, \citenamefont {Lunczer}, \citenamefont {Schlereth},
  \citenamefont {Scharf}, \citenamefont {Hankiewicz}, \citenamefont
  {Molenkamp}, \citenamefont {Halperin},\ and\ \citenamefont
  {Yacoby}}]{2019_Ren}%
  \BibitemOpen
  \bibfield  {author} {\bibinfo {author} {\bibfnamefont {H.}~\bibnamefont
  {Ren}}, \bibinfo {author} {\bibfnamefont {F.}~\bibnamefont {Pientka}},
  \bibinfo {author} {\bibfnamefont {S.}~\bibnamefont {Hart}}, \bibinfo {author}
  {\bibfnamefont {A.~T.}\ \bibnamefont {Pierce}}, \bibinfo {author}
  {\bibfnamefont {M.}~\bibnamefont {Kosowsky}}, \bibinfo {author}
  {\bibfnamefont {L.}~\bibnamefont {Lunczer}}, \bibinfo {author} {\bibfnamefont
  {R.}~\bibnamefont {Schlereth}}, \bibinfo {author} {\bibfnamefont
  {B.}~\bibnamefont {Scharf}}, \bibinfo {author} {\bibfnamefont {E.~M.}\
  \bibnamefont {Hankiewicz}}, \bibinfo {author} {\bibfnamefont {L.~W.}\
  \bibnamefont {Molenkamp}}, \bibinfo {author} {\bibfnamefont {B.~I.}\
  \bibnamefont {Halperin}},\ and\ \bibinfo {author} {\bibfnamefont
  {A.}~\bibnamefont {Yacoby}},\ }\bibfield  {title} {\bibinfo {title}
  {Topological superconductivity in a phase-controlled josephson junction},\
  }\href {https://doi.org/10.1038/s41586-019-1148-9} {\bibfield  {journal}
  {\bibinfo  {journal} {Nature}\ }\textbf {\bibinfo {volume} {569}},\ \bibinfo
  {pages} {93} (\bibinfo {year} {2019})}\BibitemShut {NoStop}%
\bibitem [{\citenamefont {Pientka}\ \emph {et~al.}(2017)\citenamefont
  {Pientka}, \citenamefont {Keselman}, \citenamefont {Berg}, \citenamefont
  {Yacoby}, \citenamefont {Stern},\ and\ \citenamefont
  {Halperin}}]{2017_Pientka}%
  \BibitemOpen
  \bibfield  {author} {\bibinfo {author} {\bibfnamefont {F.}~\bibnamefont
  {Pientka}}, \bibinfo {author} {\bibfnamefont {A.}~\bibnamefont {Keselman}},
  \bibinfo {author} {\bibfnamefont {E.}~\bibnamefont {Berg}}, \bibinfo {author}
  {\bibfnamefont {A.}~\bibnamefont {Yacoby}}, \bibinfo {author} {\bibfnamefont
  {A.}~\bibnamefont {Stern}},\ and\ \bibinfo {author} {\bibfnamefont {B.~I.}\
  \bibnamefont {Halperin}},\ }\bibfield  {title} {\bibinfo {title} {Topological
  superconductivity in a planar josephson junction},\ }\href
  {https://doi.org/10.1103/physrevx.7.021032} {\bibfield  {journal} {\bibinfo
  {journal} {Physical Review X}\ }\textbf {\bibinfo {volume} {7}},\ \bibinfo
  {pages} {021032} (\bibinfo {year} {2017})}\BibitemShut {NoStop}%
\bibitem [{\citenamefont {Pekerten}\ \emph {et~al.}(2022)\citenamefont
  {Pekerten}, \citenamefont {Pakizer}, \citenamefont {Hawn},\ and\
  \citenamefont {Matos-Abiague}}]{2022_Pekerten}%
  \BibitemOpen
  \bibfield  {author} {\bibinfo {author} {\bibfnamefont {B.}~\bibnamefont
  {Pekerten}}, \bibinfo {author} {\bibfnamefont {J.~D.}\ \bibnamefont
  {Pakizer}}, \bibinfo {author} {\bibfnamefont {B.}~\bibnamefont {Hawn}},\ and\
  \bibinfo {author} {\bibfnamefont {A.}~\bibnamefont {Matos-Abiague}},\
  }\bibfield  {title} {\bibinfo {title} {Anisotropic topological
  superconductivity in josephson junctions},\ }\href
  {https://doi.org/10.1103/physrevb.105.054504} {\bibfield  {journal} {\bibinfo
   {journal} {Physical Review B}\ }\textbf {\bibinfo {volume} {105}},\ \bibinfo
  {pages} {054504} (\bibinfo {year} {2022})}\BibitemShut {NoStop}%
\bibitem [{\citenamefont {Pakizer}\ \emph {et~al.}(2021)\citenamefont
  {Pakizer}, \citenamefont {Scharf},\ and\ \citenamefont
  {Matos-Abiague}}]{2021_Pakizer}%
  \BibitemOpen
  \bibfield  {author} {\bibinfo {author} {\bibfnamefont {J.~D.}\ \bibnamefont
  {Pakizer}}, \bibinfo {author} {\bibfnamefont {B.}~\bibnamefont {Scharf}},\
  and\ \bibinfo {author} {\bibfnamefont {A.}~\bibnamefont {Matos-Abiague}},\
  }\bibfield  {title} {\bibinfo {title} {Crystalline anisotropic topological
  superconductivity in planar josephson junctions},\ }\href
  {https://doi.org/10.1103/physrevresearch.3.013198} {\bibfield  {journal}
  {\bibinfo  {journal} {Physical Review Research}\ }\textbf {\bibinfo {volume}
  {3}},\ \bibinfo {pages} {013198} (\bibinfo {year} {2021})}\BibitemShut
  {NoStop}%
\bibitem [{\citenamefont {Pakizer}\ and\ \citenamefont
  {Matos-Abiague}(2021)}]{2021_Pakizera}%
  \BibitemOpen
  \bibfield  {author} {\bibinfo {author} {\bibfnamefont {J.~D.}\ \bibnamefont
  {Pakizer}}\ and\ \bibinfo {author} {\bibfnamefont {A.}~\bibnamefont
  {Matos-Abiague}},\ }\bibfield  {title} {\bibinfo {title} {Signatures of
  topological transitions in the spin susceptibility of josephson junctions},\
  }\href {https://doi.org/10.1103/physrevb.104.l100506} {\bibfield  {journal}
  {\bibinfo  {journal} {Physical Review B}\ }\textbf {\bibinfo {volume}
  {104}},\ \bibinfo {pages} {l100506} (\bibinfo {year} {2021})}\BibitemShut
  {NoStop}%
\bibitem [{\citenamefont {Sazgari}\ \emph {et~al.}(2020)\citenamefont
  {Sazgari}, \citenamefont {Sullivan},\ and\ \citenamefont
  {Kaya}}]{2020_Sazgari}%
  \BibitemOpen
  \bibfield  {author} {\bibinfo {author} {\bibfnamefont {V.}~\bibnamefont
  {Sazgari}}, \bibinfo {author} {\bibfnamefont {G.}~\bibnamefont {Sullivan}},\
  and\ \bibinfo {author} {\bibfnamefont {{\.{I}}.~{\.{I}}.}\ \bibnamefont
  {Kaya}},\ }\bibfield  {title} {\bibinfo {title} {Interaction-induced
  crossover between weak antilocalization and weak localization in a disordered
  {InAs}/{GaSb} double quantum well},\ }\href
  {https://doi.org/10.1103/physrevb.101.155302} {\bibfield  {journal} {\bibinfo
   {journal} {Physical Review B}\ }\textbf {\bibinfo {volume} {101}},\ \bibinfo
  {pages} {155302} (\bibinfo {year} {2020})}\BibitemShut {NoStop}%
\bibitem [{\citenamefont {Herling}\ \emph {et~al.}(2017)\citenamefont
  {Herling}, \citenamefont {Morrison}, \citenamefont {Knox}, \citenamefont
  {Zhang}, \citenamefont {Newell}, \citenamefont {Myronov}, \citenamefont
  {Linfield},\ and\ \citenamefont {Marrows}}]{2017_Herling}%
  \BibitemOpen
  \bibfield  {author} {\bibinfo {author} {\bibfnamefont {F.}~\bibnamefont
  {Herling}}, \bibinfo {author} {\bibfnamefont {C.}~\bibnamefont {Morrison}},
  \bibinfo {author} {\bibfnamefont {C.~S.}\ \bibnamefont {Knox}}, \bibinfo
  {author} {\bibfnamefont {S.}~\bibnamefont {Zhang}}, \bibinfo {author}
  {\bibfnamefont {O.}~\bibnamefont {Newell}}, \bibinfo {author} {\bibfnamefont
  {M.}~\bibnamefont {Myronov}}, \bibinfo {author} {\bibfnamefont {E.~H.}\
  \bibnamefont {Linfield}},\ and\ \bibinfo {author} {\bibfnamefont {C.~H.}\
  \bibnamefont {Marrows}},\ }\bibfield  {title} {\bibinfo {title} {Spin-orbit
  interaction in {InAs}/{GaSb} heterostructures quantified by weak
  antilocalization},\ }\href {https://doi.org/10.1103/physrevb.95.155307}
  {\bibfield  {journal} {\bibinfo  {journal} {Physical Review B}\ }\textbf
  {\bibinfo {volume} {95}},\ \bibinfo {pages} {155307} (\bibinfo {year}
  {2017})}\BibitemShut {NoStop}%
\bibitem [{\citenamefont {Spirito}\ \emph {et~al.}(2012)\citenamefont
  {Spirito}, \citenamefont {Gaspare}, \citenamefont {Evangelisti},
  \citenamefont {Gaspare}, \citenamefont {Giovine},\ and\ \citenamefont
  {Notargiacomo}}]{2012_Spirito}%
  \BibitemOpen
  \bibfield  {author} {\bibinfo {author} {\bibfnamefont {D.}~\bibnamefont
  {Spirito}}, \bibinfo {author} {\bibfnamefont {L.~D.}\ \bibnamefont
  {Gaspare}}, \bibinfo {author} {\bibfnamefont {F.}~\bibnamefont
  {Evangelisti}}, \bibinfo {author} {\bibfnamefont {A.~D.}\ \bibnamefont
  {Gaspare}}, \bibinfo {author} {\bibfnamefont {E.}~\bibnamefont {Giovine}},\
  and\ \bibinfo {author} {\bibfnamefont {A.}~\bibnamefont {Notargiacomo}},\
  }\bibfield  {title} {\bibinfo {title} {Weak antilocalization and spin-orbit
  interaction in a two-dimensional electron gas},\ }\href
  {https://doi.org/10.1103/physrevb.85.235314} {\bibfield  {journal} {\bibinfo
  {journal} {Physical Review B}\ }\textbf {\bibinfo {volume} {85}},\ \bibinfo
  {pages} {235314} (\bibinfo {year} {2012})}\BibitemShut {NoStop}%
\bibitem [{\citenamefont {Poole}\ \emph {et~al.}(1982)\citenamefont {Poole},
  \citenamefont {Pepper},\ and\ \citenamefont {Hughes}}]{1982_Poole}%
  \BibitemOpen
  \bibfield  {author} {\bibinfo {author} {\bibfnamefont {D.~A.}\ \bibnamefont
  {Poole}}, \bibinfo {author} {\bibfnamefont {M.}~\bibnamefont {Pepper}},\ and\
  \bibinfo {author} {\bibfnamefont {A.}~\bibnamefont {Hughes}},\ }\bibfield
  {title} {\bibinfo {title} {Spin-orbit coupling and weak localisation in the
  2d inversion layer of indium phosphide},\ }\href
  {https://doi.org/10.1088/0022-3719/15/32/005} {\bibfield  {journal} {\bibinfo
   {journal} {Journal of Physics C: Solid State Physics}\ }\textbf {\bibinfo
  {volume} {15}},\ \bibinfo {pages} {L1137} (\bibinfo {year}
  {1982})}\BibitemShut {NoStop}%
\bibitem [{\citenamefont {Kawaji}\ \emph {et~al.}(1985)\citenamefont {Kawaji},
  \citenamefont {Kuboki}, \citenamefont {Shigeno}, \citenamefont {Nambu},
  \citenamefont {Wakabayashi}, \citenamefont {Yoshino},\ and\ \citenamefont
  {Sakaki}}]{1985_Kawaji}%
  \BibitemOpen
  \bibfield  {author} {\bibinfo {author} {\bibfnamefont {S.}~\bibnamefont
  {Kawaji}}, \bibinfo {author} {\bibfnamefont {K.}~\bibnamefont {Kuboki}},
  \bibinfo {author} {\bibfnamefont {H.}~\bibnamefont {Shigeno}}, \bibinfo
  {author} {\bibfnamefont {T.}~\bibnamefont {Nambu}}, \bibinfo {author}
  {\bibfnamefont {J.}~\bibnamefont {Wakabayashi}}, \bibinfo {author}
  {\bibfnamefont {J.}~\bibnamefont {Yoshino}},\ and\ \bibinfo {author}
  {\bibfnamefont {H.}~\bibnamefont {Sakaki}},\ }\bibfield  {title} {\bibinfo
  {title} {Inelastic scattering and spin-orbit scattering in 2d systems of
  {GaAs}/{AlGaAs} heterostructures},\ }in\ \href
  {https://doi.org/10.1007/978-1-4615-7682-2_91} {\emph {\bibinfo {booktitle}
  {Proceedings of the 17th International Conference on the Physics of
  Semiconductors}}}\ (\bibinfo  {publisher} {Springer New York},\ \bibinfo
  {year} {1985})\ pp.\ \bibinfo {pages} {413--416}\BibitemShut {NoStop}%
\bibitem [{\citenamefont {Khmelnitskii}(1984)}]{1984_Khmelnitskii}%
  \BibitemOpen
  \bibfield  {author} {\bibinfo {author} {\bibfnamefont {D.}~\bibnamefont
  {Khmelnitskii}},\ }\bibfield  {title} {\bibinfo {title} {Localization and
  coherent scattering of electrons},\ }\href
  {https://doi.org/10.1016/0378-4363(84)90169-4} {\bibfield  {journal}
  {\bibinfo  {journal} {Physica B+C}\ }\textbf {\bibinfo {volume} {126}},\
  \bibinfo {pages} {235} (\bibinfo {year} {1984})}\BibitemShut {NoStop}%
\bibitem [{\citenamefont {Chakravarty}\ and\ \citenamefont
  {Schmid}(1986)}]{1986_Chakravarty}%
  \BibitemOpen
  \bibfield  {author} {\bibinfo {author} {\bibfnamefont {S.}~\bibnamefont
  {Chakravarty}}\ and\ \bibinfo {author} {\bibfnamefont {A.}~\bibnamefont
  {Schmid}},\ }\bibfield  {title} {\bibinfo {title} {Weak localization: The
  quasiclassical theory of electrons in a random potential},\ }\href
  {https://doi.org/10.1016/0370-1573(86)90027-x} {\bibfield  {journal}
  {\bibinfo  {journal} {Physics Reports}\ }\textbf {\bibinfo {volume} {140}},\
  \bibinfo {pages} {193} (\bibinfo {year} {1986})}\BibitemShut {NoStop}%
\bibitem [{\citenamefont {Koga}\ \emph {et~al.}(2002)\citenamefont {Koga},
  \citenamefont {Nitta}, \citenamefont {Akazaki},\ and\ \citenamefont
  {Takayanagi}}]{2002_Koga}%
  \BibitemOpen
  \bibfield  {author} {\bibinfo {author} {\bibfnamefont {T.}~\bibnamefont
  {Koga}}, \bibinfo {author} {\bibfnamefont {J.}~\bibnamefont {Nitta}},
  \bibinfo {author} {\bibfnamefont {T.}~\bibnamefont {Akazaki}},\ and\ \bibinfo
  {author} {\bibfnamefont {H.}~\bibnamefont {Takayanagi}},\ }\bibfield  {title}
  {\bibinfo {title} {Rashba spin-orbit coupling probed by the weak
  antilocalization analysis {inInAlAs}/{InGaAs}/{InAlAsQuantum} wells as a
  function of quantum well asymmetry},\ }\href
  {https://doi.org/10.1103/physrevlett.89.046801} {\bibfield  {journal}
  {\bibinfo  {journal} {Physical Review Letters}\ }\textbf {\bibinfo {volume}
  {89}},\ \bibinfo {pages} {046801} (\bibinfo {year} {2002})}\BibitemShut
  {NoStop}%
\bibitem [{\citenamefont {Miller}\ \emph {et~al.}(2003)\citenamefont {Miller},
  \citenamefont {Zumbühl}, \citenamefont {Marcus}, \citenamefont
  {Lyanda-Geller}, \citenamefont {Goldhaber-Gordon}, \citenamefont {Campman},\
  and\ \citenamefont {Gossard}}]{2003_Miller}%
  \BibitemOpen
  \bibfield  {author} {\bibinfo {author} {\bibfnamefont {J.~B.}\ \bibnamefont
  {Miller}}, \bibinfo {author} {\bibfnamefont {D.~M.}\ \bibnamefont
  {Zumbühl}}, \bibinfo {author} {\bibfnamefont {C.~M.}\ \bibnamefont
  {Marcus}}, \bibinfo {author} {\bibfnamefont {Y.~B.}\ \bibnamefont
  {Lyanda-Geller}}, \bibinfo {author} {\bibfnamefont {D.}~\bibnamefont
  {Goldhaber-Gordon}}, \bibinfo {author} {\bibfnamefont {K.}~\bibnamefont
  {Campman}},\ and\ \bibinfo {author} {\bibfnamefont {A.~C.}\ \bibnamefont
  {Gossard}},\ }\bibfield  {title} {\bibinfo {title} {Gate-controlled
  spin-orbit quantum interference effects in lateral transport},\ }\href
  {https://doi.org/10.1103/physrevlett.90.076807} {\bibfield  {journal}
  {\bibinfo  {journal} {Physical Review Letters}\ }\textbf {\bibinfo {volume}
  {90}},\ \bibinfo {pages} {076807} (\bibinfo {year} {2003})}\BibitemShut
  {NoStop}%
\bibitem [{\citenamefont {Kallaher}\ \emph {et~al.}(2010)\citenamefont
  {Kallaher}, \citenamefont {Heremans}, \citenamefont {Goel}, \citenamefont
  {Chung},\ and\ \citenamefont {Santos}}]{2010_Kallaher}%
  \BibitemOpen
  \bibfield  {author} {\bibinfo {author} {\bibfnamefont {R.~L.}\ \bibnamefont
  {Kallaher}}, \bibinfo {author} {\bibfnamefont {J.~J.}\ \bibnamefont
  {Heremans}}, \bibinfo {author} {\bibfnamefont {N.}~\bibnamefont {Goel}},
  \bibinfo {author} {\bibfnamefont {S.~J.}\ \bibnamefont {Chung}},\ and\
  \bibinfo {author} {\bibfnamefont {M.~B.}\ \bibnamefont {Santos}},\ }\bibfield
   {title} {\bibinfo {title} {Spin-orbit interaction determined by
  antilocalization in an {InSb} quantum well},\ }\href
  {https://doi.org/10.1103/physrevb.81.075303} {\bibfield  {journal} {\bibinfo
  {journal} {Physical Review B}\ }\textbf {\bibinfo {volume} {81}},\ \bibinfo
  {pages} {075303} (\bibinfo {year} {2010})}\BibitemShut {NoStop}%
\bibitem [{\citenamefont {Faniel}\ \emph {et~al.}(2011)\citenamefont {Faniel},
  \citenamefont {Matsuura}, \citenamefont {Mineshige}, \citenamefont {Sekine},\
  and\ \citenamefont {Koga}}]{2011_Faniel}%
  \BibitemOpen
  \bibfield  {author} {\bibinfo {author} {\bibfnamefont {S.}~\bibnamefont
  {Faniel}}, \bibinfo {author} {\bibfnamefont {T.}~\bibnamefont {Matsuura}},
  \bibinfo {author} {\bibfnamefont {S.}~\bibnamefont {Mineshige}}, \bibinfo
  {author} {\bibfnamefont {Y.}~\bibnamefont {Sekine}},\ and\ \bibinfo {author}
  {\bibfnamefont {T.}~\bibnamefont {Koga}},\ }\bibfield  {title} {\bibinfo
  {title} {Determination of spin-orbit coefficients in semiconductor quantum
  wells},\ }\href {https://doi.org/10.1103/physrevb.83.115309} {\bibfield
  {journal} {\bibinfo  {journal} {Physical Review B}\ }\textbf {\bibinfo
  {volume} {83}},\ \bibinfo {pages} {115309} (\bibinfo {year}
  {2011})}\BibitemShut {NoStop}%
\bibitem [{\citenamefont {Yoshizumi}\ \emph {et~al.}(2016)\citenamefont
  {Yoshizumi}, \citenamefont {Sasaki}, \citenamefont {Kohda},\ and\
  \citenamefont {Nitta}}]{2016_Yoshizumi}%
  \BibitemOpen
  \bibfield  {author} {\bibinfo {author} {\bibfnamefont {K.}~\bibnamefont
  {Yoshizumi}}, \bibinfo {author} {\bibfnamefont {A.}~\bibnamefont {Sasaki}},
  \bibinfo {author} {\bibfnamefont {M.}~\bibnamefont {Kohda}},\ and\ \bibinfo
  {author} {\bibfnamefont {J.}~\bibnamefont {Nitta}},\ }\bibfield  {title}
  {\bibinfo {title} {Gate-controlled switching between persistent and inverse
  persistent spin helix states},\ }\href {https://doi.org/10.1063/1.4944931}
  {\bibfield  {journal} {\bibinfo  {journal} {Applied Physics Letters}\
  }\textbf {\bibinfo {volume} {108}},\ \bibinfo {pages} {132402} (\bibinfo
  {year} {2016})}\BibitemShut {NoStop}%
\bibitem [{\citenamefont {Nishimura}\ \emph {et~al.}(2021)\citenamefont
  {Nishimura}, \citenamefont {Yoshizumi}, \citenamefont {Saito}, \citenamefont
  {Iizasa}, \citenamefont {Nitta},\ and\ \citenamefont
  {Kohda}}]{2021_Nishimura}%
  \BibitemOpen
  \bibfield  {author} {\bibinfo {author} {\bibfnamefont {T.}~\bibnamefont
  {Nishimura}}, \bibinfo {author} {\bibfnamefont {K.}~\bibnamefont
  {Yoshizumi}}, \bibinfo {author} {\bibfnamefont {T.}~\bibnamefont {Saito}},
  \bibinfo {author} {\bibfnamefont {D.}~\bibnamefont {Iizasa}}, \bibinfo
  {author} {\bibfnamefont {J.}~\bibnamefont {Nitta}},\ and\ \bibinfo {author}
  {\bibfnamefont {M.}~\bibnamefont {Kohda}},\ }\bibfield  {title} {\bibinfo
  {title} {Full spin-orbit coefficient in {III}-v semiconductor wires based on
  the anisotropy of weak localization under in-plane magnetic field},\ }\href
  {https://doi.org/10.1103/physrevb.103.094412} {\bibfield  {journal} {\bibinfo
   {journal} {Physical Review B}\ }\textbf {\bibinfo {volume} {103}},\ \bibinfo
  {pages} {094412} (\bibinfo {year} {2021})}\BibitemShut {NoStop}%
\bibitem [{\citenamefont {Chen}\ \emph {et~al.}(1993)\citenamefont {Chen},
  \citenamefont {Han}, \citenamefont {Huang}, \citenamefont {Datta},\ and\
  \citenamefont {Janes}}]{1993_Chen}%
  \BibitemOpen
  \bibfield  {author} {\bibinfo {author} {\bibfnamefont {G.~L.}\ \bibnamefont
  {Chen}}, \bibinfo {author} {\bibfnamefont {J.}~\bibnamefont {Han}}, \bibinfo
  {author} {\bibfnamefont {T.~T.}\ \bibnamefont {Huang}}, \bibinfo {author}
  {\bibfnamefont {S.}~\bibnamefont {Datta}},\ and\ \bibinfo {author}
  {\bibfnamefont {D.~B.}\ \bibnamefont {Janes}},\ }\bibfield  {title} {\bibinfo
  {title} {Observation of the interfacial-field-induced weak antilocalization
  in {InAs} quantum structures},\ }\href
  {https://doi.org/10.1103/physrevb.47.4084} {\bibfield  {journal} {\bibinfo
  {journal} {Physical Review B}\ }\textbf {\bibinfo {volume} {47}},\ \bibinfo
  {pages} {4084} (\bibinfo {year} {1993})}\BibitemShut {NoStop}%
\bibitem [{\citenamefont {Schierholz}\ \emph {et~al.}(2004)\citenamefont
  {Schierholz}, \citenamefont {Matsuyama}, \citenamefont {Merkt},\ and\
  \citenamefont {Meier}}]{2004_Schierholz}%
  \BibitemOpen
  \bibfield  {author} {\bibinfo {author} {\bibfnamefont {C.}~\bibnamefont
  {Schierholz}}, \bibinfo {author} {\bibfnamefont {T.}~\bibnamefont
  {Matsuyama}}, \bibinfo {author} {\bibfnamefont {U.}~\bibnamefont {Merkt}},\
  and\ \bibinfo {author} {\bibfnamefont {G.}~\bibnamefont {Meier}},\ }\bibfield
   {title} {\bibinfo {title} {Weak localization and spin splitting in inversion
  layers on p-type inas},\ }\href {https://doi.org/10.1103/physrevb.70.233311}
  {\bibfield  {journal} {\bibinfo  {journal} {Physical Review B}\ }\textbf
  {\bibinfo {volume} {70}},\ \bibinfo {pages} {233311} (\bibinfo {year}
  {2004})}\BibitemShut {NoStop}%
\bibitem [{\citenamefont {Wickramasinghe}\ \emph {et~al.}(2018)\citenamefont
  {Wickramasinghe}, \citenamefont {Mayer}, \citenamefont {Yuan}, \citenamefont
  {Nguyen}, \citenamefont {Jiao}, \citenamefont {Manucharyan},\ and\
  \citenamefont {Shabani}}]{2018_Wickramasinghe}%
  \BibitemOpen
  \bibfield  {author} {\bibinfo {author} {\bibfnamefont {K.~S.}\ \bibnamefont
  {Wickramasinghe}}, \bibinfo {author} {\bibfnamefont {W.}~\bibnamefont
  {Mayer}}, \bibinfo {author} {\bibfnamefont {J.}~\bibnamefont {Yuan}},
  \bibinfo {author} {\bibfnamefont {T.}~\bibnamefont {Nguyen}}, \bibinfo
  {author} {\bibfnamefont {L.}~\bibnamefont {Jiao}}, \bibinfo {author}
  {\bibfnamefont {V.}~\bibnamefont {Manucharyan}},\ and\ \bibinfo {author}
  {\bibfnamefont {J.}~\bibnamefont {Shabani}},\ }\bibfield  {title} {\bibinfo
  {title} {Transport properties of near surface {InAs} two-dimensional
  heterostructures},\ }\href {https://doi.org/10.1063/1.5050413} {\bibfield
  {journal} {\bibinfo  {journal} {Applied Physics Letters}\ }\textbf {\bibinfo
  {volume} {113}},\ \bibinfo {pages} {262104} (\bibinfo {year}
  {2018})}\BibitemShut {NoStop}%
\bibitem [{\citenamefont {Weigele}(2017)}]{2017_Weigele_thesis}%
  \BibitemOpen
  \bibfield  {author} {\bibinfo {author} {\bibfnamefont {P.~J.}\ \bibnamefont
  {Weigele}},\ }\emph {\bibinfo {title} {Stretching and breaking symmetry of
  the persistent spin helix in quantum transport}},\ \href
  {https://doi.org/10.5451/UNIBAS-006782930} {Ph.D. thesis} (\bibinfo {year}
  {2017})\BibitemShut {NoStop}%
\bibitem [{\citenamefont {Hikami}\ \emph {et~al.}(1980)\citenamefont {Hikami},
  \citenamefont {Larkin},\ and\ \citenamefont {Nagaoka}}]{1980_Hikami}%
  \BibitemOpen
  \bibfield  {author} {\bibinfo {author} {\bibfnamefont {S.}~\bibnamefont
  {Hikami}}, \bibinfo {author} {\bibfnamefont {A.~I.}\ \bibnamefont {Larkin}},\
  and\ \bibinfo {author} {\bibfnamefont {Y.}~\bibnamefont {Nagaoka}},\
  }\bibfield  {title} {\bibinfo {title} {Spin-orbit interaction and
  magnetoresistance in the two dimensional random system},\ }\href
  {https://doi.org/10.1143/ptp.63.707} {\bibfield  {journal} {\bibinfo
  {journal} {Progress of Theoretical Physics}\ }\textbf {\bibinfo {volume}
  {63}},\ \bibinfo {pages} {707} (\bibinfo {year} {1980})}\BibitemShut
  {NoStop}%
\bibitem [{\citenamefont {{Iordanskii}}\ \emph {et~al.}(1994)\citenamefont
  {{Iordanskii}}, \citenamefont {{Lyanda-Geller}},\ and\ \citenamefont
  {{Pikus}}}]{1994_Iordanskii}%
  \BibitemOpen
  \bibfield  {author} {\bibinfo {author} {\bibfnamefont {S.~V.}\ \bibnamefont
  {{Iordanskii}}}, \bibinfo {author} {\bibfnamefont {Y.~B.}\ \bibnamefont
  {{Lyanda-Geller}}},\ and\ \bibinfo {author} {\bibfnamefont {G.~E.}\
  \bibnamefont {{Pikus}}},\ }\bibfield  {title} {\bibinfo {title} {{Weak
  localization in quantum wells with spin-orbit interaction}},\ }\href@noop {}
  {\bibfield  {journal} {\bibinfo  {journal} {ZhETF Pisma Redaktsiiu}\ }\textbf
  {\bibinfo {volume} {60}},\ \bibinfo {pages} {199} (\bibinfo {year}
  {1994})}\BibitemShut {NoStop}%
\bibitem [{\citenamefont {Knap}\ \emph {et~al.}(1996)\citenamefont {Knap},
  \citenamefont {Skierbiszewski}, \citenamefont {Zduniak}, \citenamefont
  {Litwin-Staszewska}, \citenamefont {Bertho}, \citenamefont {Kobbi},
  \citenamefont {Robert}, \citenamefont {Pikus}, \citenamefont {Pikus},
  \citenamefont {Iordanskii}, \citenamefont {Mosser}, \citenamefont
  {Zekentes},\ and\ \citenamefont {Lyanda-Geller}}]{1996_Knap}%
  \BibitemOpen
  \bibfield  {author} {\bibinfo {author} {\bibfnamefont {W.}~\bibnamefont
  {Knap}}, \bibinfo {author} {\bibfnamefont {C.}~\bibnamefont
  {Skierbiszewski}}, \bibinfo {author} {\bibfnamefont {A.}~\bibnamefont
  {Zduniak}}, \bibinfo {author} {\bibfnamefont {E.}~\bibnamefont
  {Litwin-Staszewska}}, \bibinfo {author} {\bibfnamefont {D.}~\bibnamefont
  {Bertho}}, \bibinfo {author} {\bibfnamefont {F.}~\bibnamefont {Kobbi}},
  \bibinfo {author} {\bibfnamefont {J.~L.}\ \bibnamefont {Robert}}, \bibinfo
  {author} {\bibfnamefont {G.~E.}\ \bibnamefont {Pikus}}, \bibinfo {author}
  {\bibfnamefont {F.~G.}\ \bibnamefont {Pikus}}, \bibinfo {author}
  {\bibfnamefont {S.~V.}\ \bibnamefont {Iordanskii}}, \bibinfo {author}
  {\bibfnamefont {V.}~\bibnamefont {Mosser}}, \bibinfo {author} {\bibfnamefont
  {K.}~\bibnamefont {Zekentes}},\ and\ \bibinfo {author} {\bibfnamefont
  {Y.~B.}\ \bibnamefont {Lyanda-Geller}},\ }\bibfield  {title} {\bibinfo
  {title} {Weak antilocalization and spin precession in quantum wells},\ }\href
  {https://doi.org/10.1103/physrevb.53.3912} {\bibfield  {journal} {\bibinfo
  {journal} {Physical Review B}\ }\textbf {\bibinfo {volume} {53}},\ \bibinfo
  {pages} {3912} (\bibinfo {year} {1996})}\BibitemShut {NoStop}%
\bibitem [{\citenamefont {Pikus}\ and\ \citenamefont
  {Pikus}(1995)}]{1995_Pikus}%
  \BibitemOpen
  \bibfield  {author} {\bibinfo {author} {\bibfnamefont {F.~G.}\ \bibnamefont
  {Pikus}}\ and\ \bibinfo {author} {\bibfnamefont {G.~E.}\ \bibnamefont
  {Pikus}},\ }\bibfield  {title} {\bibinfo {title} {Conduction-band spin
  splitting and negative magnetoresistance in a3b5 heterostructures},\ }\href
  {https://doi.org/10.1103/physrevb.51.16928} {\bibfield  {journal} {\bibinfo
  {journal} {Physical Review B}\ }\textbf {\bibinfo {volume} {51}},\ \bibinfo
  {pages} {16928} (\bibinfo {year} {1995})}\BibitemShut {NoStop}%
\bibitem [{\citenamefont {Kawabata}(1984)}]{1984_Kawabata}%
  \BibitemOpen
  \bibfield  {author} {\bibinfo {author} {\bibfnamefont {A.}~\bibnamefont
  {Kawabata}},\ }\bibfield  {title} {\bibinfo {title} {On the field dependence
  of magnetoresistance in two-dimensional systems},\ }\href
  {https://doi.org/10.1143/jpsj.53.3540} {\bibfield  {journal} {\bibinfo
  {journal} {Journal of the Physical Society of Japan}\ }\textbf {\bibinfo
  {volume} {53}},\ \bibinfo {pages} {3540} (\bibinfo {year}
  {1984})}\BibitemShut {NoStop}%
\bibitem [{\citenamefont {Wittmann}\ and\ \citenamefont
  {Schmid}(1987)}]{1987_Wittmann}%
  \BibitemOpen
  \bibfield  {author} {\bibinfo {author} {\bibfnamefont {H.-P.}\ \bibnamefont
  {Wittmann}}\ and\ \bibinfo {author} {\bibfnamefont {A.}~\bibnamefont
  {Schmid}},\ }\bibfield  {title} {\bibinfo {title} {Anomalous
  magnetoconductance beyond the diffusion limit},\ }\href
  {https://doi.org/10.1007/bf00681627} {\bibfield  {journal} {\bibinfo
  {journal} {Journal of Low Temperature Physics}\ }\textbf {\bibinfo {volume}
  {69}},\ \bibinfo {pages} {131} (\bibinfo {year} {1987})}\BibitemShut
  {NoStop}%
\bibitem [{\citenamefont {Dyakonov}(1994)}]{1994_Dyakonov}%
  \BibitemOpen
  \bibfield  {author} {\bibinfo {author} {\bibfnamefont {M.}~\bibnamefont
  {Dyakonov}},\ }\bibfield  {title} {\bibinfo {title} {Magnetoconductance due
  to weak localization beyond the diffusion approximation: The high-field
  limit},\ }\href {https://doi.org/10.1016/0038-1098(94)90459-6} {\bibfield
  {journal} {\bibinfo  {journal} {Solid State Communications}\ }\textbf
  {\bibinfo {volume} {92}},\ \bibinfo {pages} {711} (\bibinfo {year}
  {1994})}\BibitemShut {NoStop}%
\bibitem [{\citenamefont {Zduniak}\ \emph {et~al.}(1997)\citenamefont
  {Zduniak}, \citenamefont {Dyakonov},\ and\ \citenamefont
  {Knap}}]{1997_Zduniak}%
  \BibitemOpen
  \bibfield  {author} {\bibinfo {author} {\bibfnamefont {A.}~\bibnamefont
  {Zduniak}}, \bibinfo {author} {\bibfnamefont {M.~I.}\ \bibnamefont
  {Dyakonov}},\ and\ \bibinfo {author} {\bibfnamefont {W.}~\bibnamefont
  {Knap}},\ }\bibfield  {title} {\bibinfo {title} {Universal behavior of
  magnetoconductance due to weak localization in two dimensions},\ }\href
  {https://doi.org/10.1103/physrevb.56.1996} {\bibfield  {journal} {\bibinfo
  {journal} {Physical Review B}\ }\textbf {\bibinfo {volume} {56}},\ \bibinfo
  {pages} {1996} (\bibinfo {year} {1997})}\BibitemShut {NoStop}%
\bibitem [{\citenamefont {McPhail}\ \emph {et~al.}(2004)\citenamefont
  {McPhail}, \citenamefont {Yasin}, \citenamefont {Hamilton}, \citenamefont
  {Simmons}, \citenamefont {Linfield}, \citenamefont {Pepper},\ and\
  \citenamefont {Ritchie}}]{2004_McPhail}%
  \BibitemOpen
  \bibfield  {author} {\bibinfo {author} {\bibfnamefont {S.}~\bibnamefont
  {McPhail}}, \bibinfo {author} {\bibfnamefont {C.~E.}\ \bibnamefont {Yasin}},
  \bibinfo {author} {\bibfnamefont {A.~R.}\ \bibnamefont {Hamilton}}, \bibinfo
  {author} {\bibfnamefont {M.~Y.}\ \bibnamefont {Simmons}}, \bibinfo {author}
  {\bibfnamefont {E.~H.}\ \bibnamefont {Linfield}}, \bibinfo {author}
  {\bibfnamefont {M.}~\bibnamefont {Pepper}},\ and\ \bibinfo {author}
  {\bibfnamefont {D.~A.}\ \bibnamefont {Ritchie}},\ }\bibfield  {title}
  {\bibinfo {title} {Weak localization in high-quality two-dimensional
  systems},\ }\href {https://doi.org/10.1103/physrevb.70.245311} {\bibfield
  {journal} {\bibinfo  {journal} {Physical Review B}\ }\textbf {\bibinfo
  {volume} {70}},\ \bibinfo {pages} {245311} (\bibinfo {year}
  {2004})}\BibitemShut {NoStop}%
\bibitem [{\citenamefont {Golub}(2005)}]{2005_Golub}%
  \BibitemOpen
  \bibfield  {author} {\bibinfo {author} {\bibfnamefont {L.~E.}\ \bibnamefont
  {Golub}},\ }\bibfield  {title} {\bibinfo {title} {Weak antilocalization in
  high-mobility two-dimensional systems},\ }\href
  {https://doi.org/10.1103/physrevb.71.235310} {\bibfield  {journal} {\bibinfo
  {journal} {Physical Review B}\ }\textbf {\bibinfo {volume} {71}},\ \bibinfo
  {pages} {235310} (\bibinfo {year} {2005})}\BibitemShut {NoStop}%
\bibitem [{\citenamefont {Sawada}\ and\ \citenamefont
  {Koga}(2017)}]{2017_Sawada}%
  \BibitemOpen
  \bibfield  {author} {\bibinfo {author} {\bibfnamefont {A.}~\bibnamefont
  {Sawada}}\ and\ \bibinfo {author} {\bibfnamefont {T.}~\bibnamefont {Koga}},\
  }\bibfield  {title} {\bibinfo {title} {Universal modeling of weak
  antilocalization corrections in quasi-two-dimensional electron systems using
  predetermined return orbitals},\ }\href
  {https://doi.org/10.1103/physreve.95.023309} {\bibfield  {journal} {\bibinfo
  {journal} {Physical Review E}\ }\textbf {\bibinfo {volume} {95}},\ \bibinfo
  {pages} {023309} (\bibinfo {year} {2017})}\BibitemShut {NoStop}%
\bibitem [{\citenamefont {Marinescu}\ \emph {et~al.}(2019)\citenamefont
  {Marinescu}, \citenamefont {Weigele}, \citenamefont {Zumbühl},\ and\
  \citenamefont {Egues}}]{2019_Marinescu}%
  \BibitemOpen
  \bibfield  {author} {\bibinfo {author} {\bibfnamefont {D.}~\bibnamefont
  {Marinescu}}, \bibinfo {author} {\bibfnamefont {P.~J.}\ \bibnamefont
  {Weigele}}, \bibinfo {author} {\bibfnamefont {D.~M.}\ \bibnamefont
  {Zumbühl}},\ and\ \bibinfo {author} {\bibfnamefont {J.~C.}\ \bibnamefont
  {Egues}},\ }\bibfield  {title} {\bibinfo {title} {Closed-form weak
  localization magnetoconductivity in quantum wells with arbitrary rashba and
  dresselhaus spin-orbit interactions},\ }\href
  {https://doi.org/10.1103/physrevlett.122.156601} {\bibfield  {journal}
  {\bibinfo  {journal} {Physical Review Letters}\ }\textbf {\bibinfo {volume}
  {122}},\ \bibinfo {pages} {156601} (\bibinfo {year} {2019})}\BibitemShut
  {NoStop}%
\bibitem [{\citenamefont {Mal'shukov}\ \emph {et~al.}(1999)\citenamefont
  {Mal'shukov}, \citenamefont {Froltsov},\ and\ \citenamefont
  {Chao}}]{1999_Malshukov}%
  \BibitemOpen
  \bibfield  {author} {\bibinfo {author} {\bibfnamefont {A.~G.}\ \bibnamefont
  {Mal'shukov}}, \bibinfo {author} {\bibfnamefont {V.~A.}\ \bibnamefont
  {Froltsov}},\ and\ \bibinfo {author} {\bibfnamefont {K.~A.}\ \bibnamefont
  {Chao}},\ }\bibfield  {title} {\bibinfo {title} {Crystal anisotropy effects
  on the weak-localization magnetoresistance of a {III}-v semiconductor quantum
  well in a magnetic field parallel to interfaces},\ }\href
  {https://doi.org/10.1103/physrevb.59.5702} {\bibfield  {journal} {\bibinfo
  {journal} {Physical Review B}\ }\textbf {\bibinfo {volume} {59}},\ \bibinfo
  {pages} {5702} (\bibinfo {year} {1999})}\BibitemShut {NoStop}%
\bibitem [{\citenamefont {Hupfauer}\ \emph {et~al.}(2015)\citenamefont
  {Hupfauer}, \citenamefont {Matos-Abiague}, \citenamefont {Gmitra},
  \citenamefont {Schiller}, \citenamefont {Loher}, \citenamefont {Bougeard},
  \citenamefont {Back}, \citenamefont {Fabian},\ and\ \citenamefont
  {Weiss}}]{2015_Hupfauer}%
  \BibitemOpen
  \bibfield  {author} {\bibinfo {author} {\bibfnamefont {T.}~\bibnamefont
  {Hupfauer}}, \bibinfo {author} {\bibfnamefont {A.}~\bibnamefont
  {Matos-Abiague}}, \bibinfo {author} {\bibfnamefont {M.}~\bibnamefont
  {Gmitra}}, \bibinfo {author} {\bibfnamefont {F.}~\bibnamefont {Schiller}},
  \bibinfo {author} {\bibfnamefont {J.}~\bibnamefont {Loher}}, \bibinfo
  {author} {\bibfnamefont {D.}~\bibnamefont {Bougeard}}, \bibinfo {author}
  {\bibfnamefont {C.~H.}\ \bibnamefont {Back}}, \bibinfo {author}
  {\bibfnamefont {J.}~\bibnamefont {Fabian}},\ and\ \bibinfo {author}
  {\bibfnamefont {D.}~\bibnamefont {Weiss}},\ }\bibfield  {title} {\bibinfo
  {title} {Emergence of spin{\textendash}orbit fields in magnetotransport of
  quasi-two-dimensional iron on gallium arsenide},\ }\bibfield  {journal}
  {\bibinfo  {journal} {Nature Communications}\ }\textbf {\bibinfo {volume}
  {6}},\ \href {https://doi.org/10.1038/ncomms8374} {10.1038/ncomms8374}
  (\bibinfo {year} {2015})\BibitemShut {NoStop}%
\bibitem [{\citenamefont {Jungwirth}\ \emph {et~al.}(2014)\citenamefont
  {Jungwirth}, \citenamefont {Wunderlich}, \citenamefont {Nov{\'{a}}k},
  \citenamefont {Olejn{\'{\i}}k}, \citenamefont {Gallagher}, \citenamefont
  {Campion}, \citenamefont {Edmonds}, \citenamefont {Rushforth}, \citenamefont
  {Ferguson},\ and\ \citenamefont {N{\v{e}}mec}}]{2014_Jungwirth}%
  \BibitemOpen
  \bibfield  {author} {\bibinfo {author} {\bibfnamefont {T.}~\bibnamefont
  {Jungwirth}}, \bibinfo {author} {\bibfnamefont {J.}~\bibnamefont
  {Wunderlich}}, \bibinfo {author} {\bibfnamefont {V.}~\bibnamefont
  {Nov{\'{a}}k}}, \bibinfo {author} {\bibfnamefont {K.}~\bibnamefont
  {Olejn{\'{\i}}k}}, \bibinfo {author} {\bibfnamefont {B.}~\bibnamefont
  {Gallagher}}, \bibinfo {author} {\bibfnamefont {R.}~\bibnamefont {Campion}},
  \bibinfo {author} {\bibfnamefont {K.}~\bibnamefont {Edmonds}}, \bibinfo
  {author} {\bibfnamefont {A.}~\bibnamefont {Rushforth}}, \bibinfo {author}
  {\bibfnamefont {A.}~\bibnamefont {Ferguson}},\ and\ \bibinfo {author}
  {\bibfnamefont {P.}~\bibnamefont {N{\v{e}}mec}},\ }\bibfield  {title}
  {\bibinfo {title} {Spin-dependent phenomena and device concepts explored in
  (ga,mn)as},\ }\href {https://doi.org/10.1103/revmodphys.86.855} {\bibfield
  {journal} {\bibinfo  {journal} {Reviews of Modern Physics}\ }\textbf
  {\bibinfo {volume} {86}},\ \bibinfo {pages} {855} (\bibinfo {year}
  {2014})}\BibitemShut {NoStop}%
\bibitem [{\citenamefont {Löhr}\ \emph {et~al.}(2003)\citenamefont {Löhr},
  \citenamefont {Mendach}, \citenamefont {Vonau}, \citenamefont {Heyn},\ and\
  \citenamefont {Hansen}}]{2003_Loehr}%
  \BibitemOpen
  \bibfield  {author} {\bibinfo {author} {\bibfnamefont {S.}~\bibnamefont
  {Löhr}}, \bibinfo {author} {\bibfnamefont {S.}~\bibnamefont {Mendach}},
  \bibinfo {author} {\bibfnamefont {T.}~\bibnamefont {Vonau}}, \bibinfo
  {author} {\bibfnamefont {C.}~\bibnamefont {Heyn}},\ and\ \bibinfo {author}
  {\bibfnamefont {W.}~\bibnamefont {Hansen}},\ }\bibfield  {title} {\bibinfo
  {title} {Highly anisotropic electron transport in shallow {InGaAs}
  heterostructures},\ }\href {https://doi.org/10.1103/physrevb.67.045309}
  {\bibfield  {journal} {\bibinfo  {journal} {Physical Review B}\ }\textbf
  {\bibinfo {volume} {67}},\ \bibinfo {pages} {045309} (\bibinfo {year}
  {2003})}\BibitemShut {NoStop}%
\bibitem [{\citenamefont {{\v{Z}}uti{\'{c}}}\ \emph {et~al.}(2004)\citenamefont
  {{\v{Z}}uti{\'{c}}}, \citenamefont {Fabian},\ and\ \citenamefont
  {Sarma}}]{2004_Zutic}%
  \BibitemOpen
  \bibfield  {author} {\bibinfo {author} {\bibfnamefont {I.}~\bibnamefont
  {{\v{Z}}uti{\'{c}}}}, \bibinfo {author} {\bibfnamefont {J.}~\bibnamefont
  {Fabian}},\ and\ \bibinfo {author} {\bibfnamefont {S.~D.}\ \bibnamefont
  {Sarma}},\ }\bibfield  {title} {\bibinfo {title} {Spintronics: Fundamentals
  and applications},\ }\href {https://doi.org/10.1103/revmodphys.76.323}
  {\bibfield  {journal} {\bibinfo  {journal} {Reviews of Modern Physics}\
  }\textbf {\bibinfo {volume} {76}},\ \bibinfo {pages} {323} (\bibinfo {year}
  {2004})}\BibitemShut {NoStop}%
\bibitem [{\citenamefont {Fabian}\ \emph {et~al.}(2007)\citenamefont {Fabian},
  \citenamefont {Matos-Abiague}, \citenamefont {Ertler}, \citenamefont
  {Stano},\ and\ \citenamefont {{\v{Z}}uti{\'{c}}}}]{2007_Fabian}%
  \BibitemOpen
  \bibfield  {author} {\bibinfo {author} {\bibfnamefont {J.}~\bibnamefont
  {Fabian}}, \bibinfo {author} {\bibfnamefont {A.}~\bibnamefont
  {Matos-Abiague}}, \bibinfo {author} {\bibfnamefont {C.}~\bibnamefont
  {Ertler}}, \bibinfo {author} {\bibfnamefont {P.}~\bibnamefont {Stano}},\ and\
  \bibinfo {author} {\bibfnamefont {I.}~\bibnamefont {{\v{Z}}uti{\'{c}}}},\
  }\bibfield  {title} {\bibinfo {title} {Semiconductor spintronics},\
  }\bibfield  {journal} {\bibinfo  {journal} {Acta Physica Slovaca. Reviews and
  Tutorials}\ }\textbf {\bibinfo {volume} {57}},\ \href
  {https://doi.org/10.2478/v10155-010-0086-8} {10.2478/v10155-010-0086-8}
  (\bibinfo {year} {2007})\BibitemShut {NoStop}%
\bibitem [{\citenamefont {Yuan}\ \emph {et~al.}(2020)\citenamefont {Yuan},
  \citenamefont {Hatefipour}, \citenamefont {Magill}, \citenamefont {Mayer},
  \citenamefont {Dartiailh}, \citenamefont {Sardashti}, \citenamefont
  {Wickramasinghe}, \citenamefont {Khodaparast}, \citenamefont {Matsuda},
  \citenamefont {Kohama}, \citenamefont {Yang}, \citenamefont {Thapa},
  \citenamefont {Stanton},\ and\ \citenamefont {Shabani}}]{2020_Yuan}%
  \BibitemOpen
  \bibfield  {author} {\bibinfo {author} {\bibfnamefont {J.}~\bibnamefont
  {Yuan}}, \bibinfo {author} {\bibfnamefont {M.}~\bibnamefont {Hatefipour}},
  \bibinfo {author} {\bibfnamefont {B.~A.}\ \bibnamefont {Magill}}, \bibinfo
  {author} {\bibfnamefont {W.}~\bibnamefont {Mayer}}, \bibinfo {author}
  {\bibfnamefont {M.~C.}\ \bibnamefont {Dartiailh}}, \bibinfo {author}
  {\bibfnamefont {K.}~\bibnamefont {Sardashti}}, \bibinfo {author}
  {\bibfnamefont {K.~S.}\ \bibnamefont {Wickramasinghe}}, \bibinfo {author}
  {\bibfnamefont {G.~A.}\ \bibnamefont {Khodaparast}}, \bibinfo {author}
  {\bibfnamefont {Y.~H.}\ \bibnamefont {Matsuda}}, \bibinfo {author}
  {\bibfnamefont {Y.}~\bibnamefont {Kohama}}, \bibinfo {author} {\bibfnamefont
  {Z.}~\bibnamefont {Yang}}, \bibinfo {author} {\bibfnamefont {S.}~\bibnamefont
  {Thapa}}, \bibinfo {author} {\bibfnamefont {C.~J.}\ \bibnamefont {Stanton}},\
  and\ \bibinfo {author} {\bibfnamefont {J.}~\bibnamefont {Shabani}},\
  }\bibfield  {title} {\bibinfo {title} {Experimental measurements of effective
  mass in near-surface {InAs} quantum wells},\ }\href
  {https://doi.org/10.1103/physrevb.101.205310} {\bibfield  {journal} {\bibinfo
   {journal} {Physical Review B}\ }\textbf {\bibinfo {volume} {101}},\ \bibinfo
  {pages} {205310} (\bibinfo {year} {2020})}\BibitemShut {NoStop}%
\bibitem [{\citenamefont {Minkov}\ \emph {et~al.}(2004)\citenamefont {Minkov},
  \citenamefont {Germanenko}, \citenamefont {Rut}, \citenamefont
  {Sherstobitov}, \citenamefont {Golub}, \citenamefont {Zvonkov},\ and\
  \citenamefont {Willander}}]{2004_Minkov}%
  \BibitemOpen
  \bibfield  {author} {\bibinfo {author} {\bibfnamefont {G.~M.}\ \bibnamefont
  {Minkov}}, \bibinfo {author} {\bibfnamefont {A.~V.}\ \bibnamefont
  {Germanenko}}, \bibinfo {author} {\bibfnamefont {O.~E.}\ \bibnamefont {Rut}},
  \bibinfo {author} {\bibfnamefont {A.~A.}\ \bibnamefont {Sherstobitov}},
  \bibinfo {author} {\bibfnamefont {L.~E.}\ \bibnamefont {Golub}}, \bibinfo
  {author} {\bibfnamefont {B.~N.}\ \bibnamefont {Zvonkov}},\ and\ \bibinfo
  {author} {\bibfnamefont {M.}~\bibnamefont {Willander}},\ }\bibfield  {title}
  {\bibinfo {title} {Weak antilocalization in quantum wells in tilted magnetic
  fields},\ }\href {https://doi.org/10.1103/physrevb.70.155323} {\bibfield
  {journal} {\bibinfo  {journal} {Physical Review B}\ }\textbf {\bibinfo
  {volume} {70}},\ \bibinfo {pages} {155323} (\bibinfo {year}
  {2004})}\BibitemShut {NoStop}%
\bibitem [{\citenamefont {Grundler}(2000)}]{2000_Grundler}%
  \BibitemOpen
  \bibfield  {author} {\bibinfo {author} {\bibfnamefont {D.}~\bibnamefont
  {Grundler}},\ }\bibfield  {title} {\bibinfo {title} {Large rashba splitting
  in inas quantum wells due to electron wave function penetration into the
  barrier layers},\ }\href {https://doi.org/10.1103/physrevlett.84.6074}
  {\bibfield  {journal} {\bibinfo  {journal} {Physical Review Letters}\
  }\textbf {\bibinfo {volume} {84}},\ \bibinfo {pages} {6074} (\bibinfo {year}
  {2000})}\BibitemShut {NoStop}%
\bibitem [{\citenamefont {Oestreich}\ \emph {et~al.}(1996)\citenamefont
  {Oestreich}, \citenamefont {Hallstein},\ and\ \citenamefont
  {Ruhle}}]{1996_Oestreich}%
  \BibitemOpen
  \bibfield  {author} {\bibinfo {author} {\bibfnamefont {M.}~\bibnamefont
  {Oestreich}}, \bibinfo {author} {\bibfnamefont {S.}~\bibnamefont
  {Hallstein}},\ and\ \bibinfo {author} {\bibfnamefont {W.}~\bibnamefont
  {Ruhle}},\ }\bibfield  {title} {\bibinfo {title} {Spin quantum beats in
  semiconductors},\ }\href {https://doi.org/10.1109/2944.571776} {\bibfield
  {journal} {\bibinfo  {journal} {{IEEE} Journal of Selected Topics in Quantum
  Electronics}\ }\textbf {\bibinfo {volume} {2}},\ \bibinfo {pages} {747}
  (\bibinfo {year} {1996})}\BibitemShut {NoStop}%
\bibitem [{\citenamefont {Trellakis}\ \emph {et~al.}(2006)\citenamefont
  {Trellakis}, \citenamefont {Zibold}, \citenamefont {Andlauer}, \citenamefont
  {Birner}, \citenamefont {Smith}, \citenamefont {Morschl},\ and\ \citenamefont
  {Vogl}}]{2006_Trellakis}%
  \BibitemOpen
  \bibfield  {author} {\bibinfo {author} {\bibfnamefont {A.}~\bibnamefont
  {Trellakis}}, \bibinfo {author} {\bibfnamefont {T.}~\bibnamefont {Zibold}},
  \bibinfo {author} {\bibfnamefont {T.}~\bibnamefont {Andlauer}}, \bibinfo
  {author} {\bibfnamefont {S.}~\bibnamefont {Birner}}, \bibinfo {author}
  {\bibfnamefont {R.~K.}\ \bibnamefont {Smith}}, \bibinfo {author}
  {\bibfnamefont {R.}~\bibnamefont {Morschl}},\ and\ \bibinfo {author}
  {\bibfnamefont {P.}~\bibnamefont {Vogl}},\ }\bibfield  {title} {\bibinfo
  {title} {The 3d nanometer device project nextnano: Concepts, methods,
  results},\ }\href {https://doi.org/10.1007/s10825-006-0005-x} {\bibfield
  {journal} {\bibinfo  {journal} {Journal of Computational Electronics}\
  }\textbf {\bibinfo {volume} {5}},\ \bibinfo {pages} {285} (\bibinfo {year}
  {2006})}\BibitemShut {NoStop}%
\bibitem [{\citenamefont {Vurgaftman}\ \emph {et~al.}(2001)\citenamefont
  {Vurgaftman}, \citenamefont {Meyer},\ and\ \citenamefont
  {Ram-Mohan}}]{2001_Vurgaftman}%
  \BibitemOpen
  \bibfield  {author} {\bibinfo {author} {\bibfnamefont {I.}~\bibnamefont
  {Vurgaftman}}, \bibinfo {author} {\bibfnamefont {J.~R.}\ \bibnamefont
  {Meyer}},\ and\ \bibinfo {author} {\bibfnamefont {L.~R.}\ \bibnamefont
  {Ram-Mohan}},\ }\bibfield  {title} {\bibinfo {title} {Band parameters for
  {III}{\textendash}v compound semiconductors and their alloys},\ }\href
  {https://doi.org/10.1063/1.1368156} {\bibfield  {journal} {\bibinfo
  {journal} {Journal of Applied Physics}\ }\textbf {\bibinfo {volume} {89}},\
  \bibinfo {pages} {5815} (\bibinfo {year} {2001})}\BibitemShut {NoStop}%
\bibitem [{\citenamefont {Winkler}(2003)}]{2003_Winkler_book}%
  \BibitemOpen
  \bibfield  {author} {\bibinfo {author} {\bibfnamefont {R.}~\bibnamefont
  {Winkler}},\ }\href {https://doi.org/10.1007/b13586} {\emph {\bibinfo {title}
  {Spin-Orbit Coupling Effects in Two-Dimensional Electron and Hole Systems}}}\
  (\bibinfo  {publisher} {Springer Berlin Heidelberg},\ \bibinfo {year}
  {2003})\BibitemShut {NoStop}%
\bibitem [{\citenamefont {{Kalevich}}\ and\ \citenamefont
  {{Korenev}}(1993)}]{1993_Kalevich}%
  \BibitemOpen
  \bibfield  {author} {\bibinfo {author} {\bibfnamefont {V.~K.}\ \bibnamefont
  {{Kalevich}}}\ and\ \bibinfo {author} {\bibfnamefont {V.~L.}\ \bibnamefont
  {{Korenev}}},\ }\bibfield  {title} {\bibinfo {title} {{Electron g-factor
  anisotropy in asymmetric GaAs/AlGaAs quantum well}},\ }\href@noop {}
  {\bibfield  {journal} {\bibinfo  {journal} {Soviet Journal of Experimental
  and Theoretical Physics Letters}\ }\textbf {\bibinfo {volume} {57}},\
  \bibinfo {pages} {571} (\bibinfo {year} {1993})}\BibitemShut {NoStop}%
\end{thebibliography}%

\clearpage
\newpage
\onecolumngrid
\setcounter{page}{1}
\setcounter{section}{0}
\setcounter{equation}{0}
\setcounter{figure}{0}
\setcounter{table}{0}
\renewcommand{\thepage}{S\arabic{page}}
\renewcommand{\thesection}{S\arabic{section}}
\renewcommand{\theequation}{\thesection.\arabic{equation}}
\renewcommand{\thefigure}{S\arabic{figure}}
\renewcommand{\thetable}{S\arabic{table}}

\section*{Supplemental Material}

% \twocolumngrid
\section{Simulations}
Here we estimate the strength of Rashba and Dresselhaus spin-orbit couplings from the solutions of Poisson-Schrodinger equations. 
The solutions are obtained by assuming a single band Kane model where the bands are decoupled through k dot p approximation. 
We use the resulting electron wavefunction which contains only the spatial part as the spin part is unresolved in these calculations. 
With the proper choice of basis the wavefunctions can be real valued and therefore obtained from the charge distribution.
In a bulk zinc blende crystal the Rashba strength is given as $\alpha = r_{41}^{6c6c} \ev{\mathcal{E}_z}$ and the Dresselhaus strength as $\gamma = b_{41}^{6c6c}$, where $r_{41}^{6c6c}$ and $b_{41}^{6c6c}$ are material-dependent parameters \cite{2003_Winkler_book} and $\mathcal{E}_z = -dV(z)/dz$ is the extrinsic electric field applied on the crystal. 

The picture is modified in a quasi-two-dimensional quantum well as the confinement leads to a higher band gap which is a result of quantization of energy levels, i.e., subbands. 
As a result, in a first order approximation, the Dresselhaus terms modifies to $\gamma = b_{52}^{6c6c}$ and $\beta = b_{52}^{6c6c} \ev{k_z^2}$ where $\ev{k_z^2}$ is the average momentum squared of the wavefunction in the $z$ direction, i.e., the confinement direction. 
The parameters $r_{41}^{6c6c}$ and $b_{52}^{6c6c}$ become subband dependent as they depend on the energy spacing between different subbbands. 
Here we modify these parameters in a first order approximation where the bands above the gap get shifted upward by the quantization energy whereas the bands below the gap get shifted downward by the same amounts. Generally, $r_{41}^{6c6c}$ and $b_{52}^{6c6c}$ are position dependent as well, however, we assume that the dependence is negligible as the composition of the ternary alloy In$_{0.81}$Ga$_{0.19}$As is close to that of InAs and therefore one expects the corresponding Kane parameters to be similar as well. 

Assuming that more than one subband of the quantum well is occupied, one can write the wavefunction as a linear combination of subband eigenfunctions, i.e., $\psi(z) = \sum_i c_i\psi_i(z)$, where $\abs{c_i}^2$ denotes the occupation number of the $i$-th subband. 
Therefore, 
\begin{equation}
    \alpha = -\sum_i r_{41,i}^{6c6c} \abs{c_i}^2 \int dz \psi_i^*(z) \big(\frac{dV(z)}{dz}\big) \psi_i(z),
\end{equation}
\begin{equation}
    \beta = -\sum_i b_{52_i}^{6c6c} \abs{c_i}^2 \int dz \psi_i^*(z) \frac{d^2\psi_i(z)}{dz^2},
\end{equation}
\begin{equation}
    \gamma = \sum_i b_{52,i}^{6c6c} \abs{c_i}^2. 
\end{equation}

\begin{figure}[ht]
    \centering
    \begin{tabular}{cc}
        \includegraphics[width=0.33\linewidth]{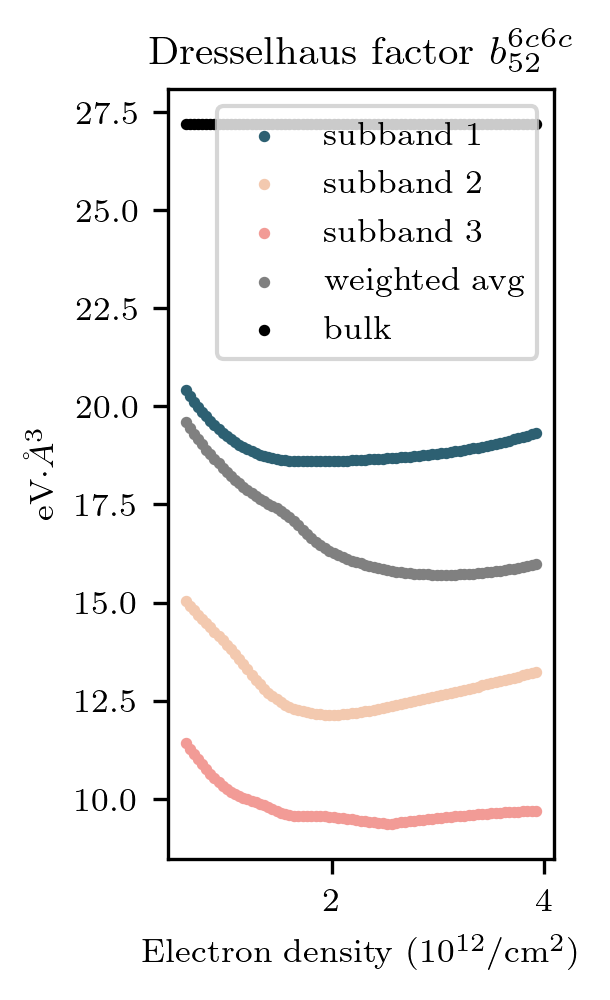} & 
        \includegraphics[width=0.33\linewidth]{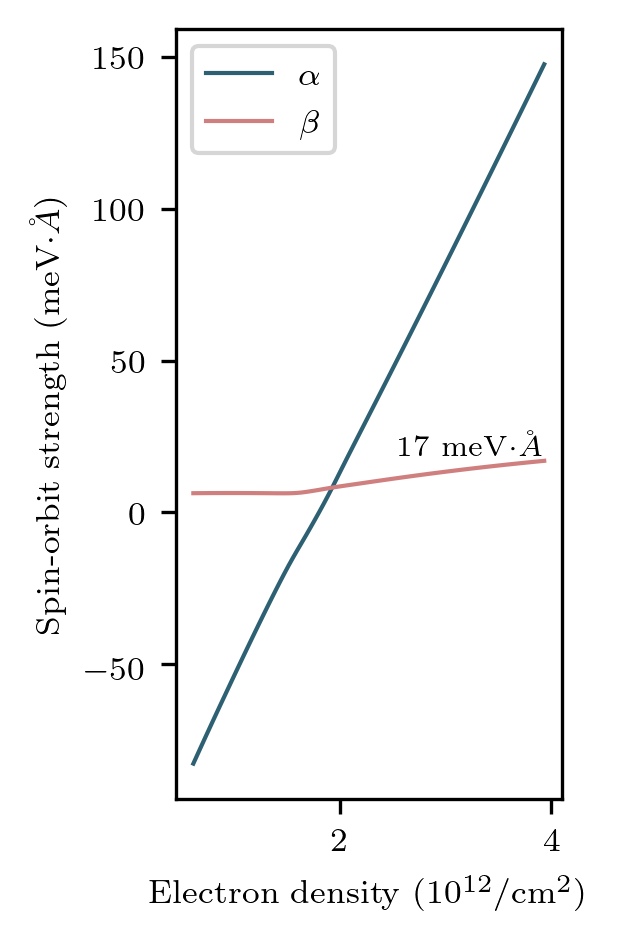}
    \end{tabular}
    
    \caption{\label{fig:sim_beta} The calculated strength of the Rashba and Dresselhaus spin-orbit coupling versus total electron density.}
\end{figure}

The calculated values of $\beta$ are plotted in Fig. \ref{fig:sim_beta} as a function of total electron density. As the occupation number of the second subband increases, $\ev{k_z^2}$ increases as well because the charge density in the second subband contains two humps instead of one hump in the first subband. The onset of second subband is at \SI{1.5e12}{cm^{-2}}. 
We use these values as initial values for the fitting procedure of the weak antilocalization measurements.

% simplifying assumptions  
% quantum well is spatially inhomogeneous and is made up of ternary alloys. 
% For instance, we define the average Rashba parameter of the quantum well as a weighted sum over the Rashba parameter of different layers where the weights are the charge density and the subband occupation. 

\section{Comparison to the ILP method}
\begin{figure}[ht]
    \centering
    \includegraphics[width=0.5\linewidth]{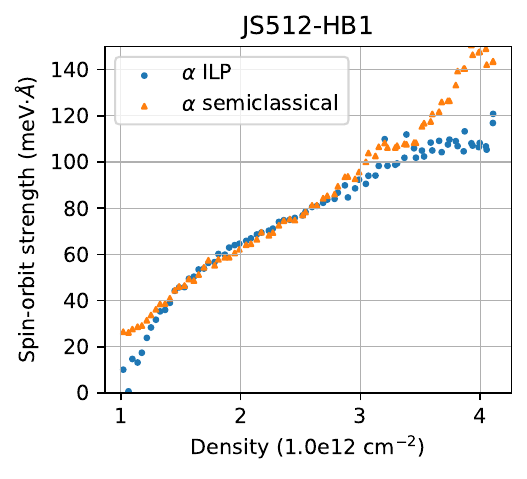} 
    \caption{\label{fig:supp_comparison} The comparison between two different theoretical models for describing the magnetoconductivity in the weak antilocalization regime. The Iordanskii, Lyanda-Geller, and Pikus (ILP) \cite{1996_Knap} method is a quantum mechanical theory based on the Landau quantization of the energy levels whereas the semiclassical method by Sawada and Koga \cite{2017_Sawada} assumes electrons obey the Boltzmann transport equation.}
\end{figure}

The weak localization model by Iordanskii, Lyanda-Geller, and Pikus (ILP) is based on the solution to the cooperon equation. The model was later modified by Knap et al. 1996 \cite{1996_Knap}. 
To compare our semiclassical approach with the ILP method, we calculate the Rashba spin-orbit strength $\alpha$ (assuming no linearized dresselhaus, i.e., $\beta = 0$) using both methods. The results are plotted in Fig.  \ref{fig:supp_comparison}. 

\section{\label{sec:aniso} Ruling out other contributions}
In the presence of an in-plane magnetic field $B_\parallel$, the Hamiltonian includes additional terms that make the electron g-factor anisotropic. 
We show that these terms are negligible compared to the Zeeman splitting $g_\parallel\mu_\mathrm{B}\vb*{B}_\parallel\cdot\vb*{\sigma}$ extracted from magnetoconductivity measurements or compared to the strength of spin-orbit coupling. 

\subsection{Coupling of $B_\parallel$ and $\mathcal{E}_z$}
Due to the reduced symmetry of the quantum well in the growth direction, one can show \cite{2003_Winkler_book}, based on the theory of invariants, that the built-in electric field $\mathcal{E}_z$ and the in-plane magnetic field $B_\parallel$ can couple in the following invariant 
\begin{equation}
    H = z_{41}^{6c6c}\mathcal{E}_z (B_y \sigma_x + B_x \sigma_y).
\end{equation}
This term modifies the electron g-factor from a scalar to a tensor which can accommodate an anisotropic Zeeman splitting. 
The parameter $z_{41}^{6c6c}$ is derived from the Kane model for the $\Gamma_6^c$ band. 
From the bulk parameters one obtains $z_{41}^{6c6c}=$ \SI{-0.328}{me\AA/T} for InAs. 
As both simulations and experiments suggest the highest achievable electric field due to the built-in potential or the gate voltage is at most of the order of $\abs{\mathcal{E}_z} = $ \SI{1e-2}{V/\AA}. 
Therefore, a magnetic field of $B_\parallel=$ \SI{0.2}{T} results in a spin splitting of \SI{0.7}{\micro eV} which is three orders of magnitude smaller than the experimental values we extracted from the weak antilocalization measurements. 

\subsection{Orbital effects}
In the presence of an in-plane magnetic field the electrons obtain an orbital angular momentum in the plane of the quantum well. This effect is pronounced when the magnetic field length $\sqrt{\hbar/eB_\parallel}$ is comparable to the width of the quantum well. For a \SI{20}{nm} quantum well, the corresponding magnetic field is $B_\parallel = $ \SI{1.3}{T}. Since we apply magnetic fields as large as \SI{0.2}{T} in our experiments, it is important to quantify the resulting orbital effects. 
It has been shown \cite{2003_Winkler_book, 1993_Kalevich} that using a symmetric gauge, the Dresselhaus term in the $\Gamma_6^c$ band gets modified, to the leading order in momentum, as 
\begin{equation}
\begin{split}
    H & = b_{41}^{6c6c} \ev{k_z^2}(-\mathfrak{k}_x\sigma_x + \mathfrak{k}_y\sigma_y) \\ 
    & + b_{41}^{6c6c}(\ev{k_z^2}\ev{z} - \ev{\{k_z^2, z\}})\frac{e}{\hbar}(B_y\sigma_x + B_x\sigma_y), 
\end{split}
\end{equation}
where $b_{41}^{6c6c}=$ \SI{27.18}{eV\AA^3} is a parameter of bulk InAs, $\mathfrak{k}$ denotes the kinetic momentum, and $\{k_z^2, z\} = (k_z^2z + zk_z^2)/2$.  
Using the Poisson-Schrodinger's solutions one can calculate the strength of the second term in the equation above which results in a \SI{6}{\micro eV} spin splitting for $B_\parallel=$ \SI{0.2}{T}. This is an upper bound since the calculation is done at high densities where the higher subbands are occupied and therefore $k_z^2$ is maximized. 
We note that this value is at least an order of magnitude smaller than the Dresselhaus strength $\beta k_\mathrm{F} = b_{41}^{6c6c}\ev{k_z^2} k_\mathrm{F}$ which is \SI{150}{\micro eV} for an electron density of \SI{1e12}{\per cm\squared}.

\section{Phenomenological model}
Here we present a simple phenomenological model that reveals how the anisotropic behavior of the conductivity with respect to the orientation of the in-plane magnetic field and its specific angular dependence is closely related to the symmetry of the spin-orbit field. 

Ignoring orbital effects, the SOC Hamiltonian can be written as,
\begin{equation}
    H_\mathrm{so}=\vb*{w}(\vb*{k}_\parallel)\cdot\boldsymbol{\sigma},
\end{equation}
where,
\begin{equation}\label{eq:sof}
    \vb*{w}(\vb*{k}_\parallel)=\begin{pmatrix}
              \alpha k_y +\beta k_x +\gamma k_x k_y^2\\
              -\alpha k_x - \beta k_y -\gamma k_y k_x^2\\
              0
              \end{pmatrix}
\end{equation}
is the spin-orbit field with $\alpha$, $\beta$, and $\gamma$ representing the Rashba and linearized and cubic Dresselhaus parameters, respectively.

There are only two preferential directions related to the in-plane anisotropy, namely the spin-orbit and in-plane magnetic field directions. Therefore, a scalar quantity such as the longitudinal conductivity can be expanded in a series of the all possible
scalars one can form with the vectors $\vb*{w}(\vb*{k}_\parallel)$ and $\vb*{B}_\parallel$. Up to second order the conductivity reads,
\begin{equation}
\label{eq:s-expand}
%\begin{split}
        \sigma \approx \sigma^0 + \ev{ a_{11}(\vb*{w}\cdot\vb*{B}_\parallel)} + \ev{a_{20}w^2} +
        \ev{a_{02}B_\parallel^2} \nonumber
        + \langle a_{22}(\vb*{w}\cdot\vb*{B}_\parallel)^2\rangle
%\end{split}
\end{equation}
where $\sigma^0$ refers to the conductivity in the absence of spin-orbit and in-plane magnetic fields. The expansion coefficients $a_{ij}=a_{ij}(\vb*{k}_\parallel)$ are SOC-independent and, therefore, isotropic in the $\vb*{k}_\parallel$-space. We have used the notation $\langle ...\rangle$ to indicate momentum average. Since the spin-orbit field is odd in momentum, the momentum average of the contributions of first order in $\vb*{w}$ vanish. The terms $\ev{ a_{20}w^2}$ and $ \ev{a_{02}B_\parallel^2}$ do not depend on the in-plane magnetic field direction and represent isotropic corrections to the conductivity. Therefore, the in-plane anisotropy of the conductivity is governed by the last contribution in Eq.~\ref{eq:s-expand}, which after substituting the spin-orbit field in Eq.~(\ref{eq:sof}) and taking into account that contributions odd in the momentum vanish after momentum average, reduces to
\begin{equation}
    \langle a_{22}(\vb*{w}\cdot\vb*{B}_\parallel)^2\rangle = \left\langle \frac{a_{22}}{2}\left[(\alpha^2+\beta^2)k_\parallel^2+\frac{\beta\gamma}{2}k_{\parallel}^4+\frac{\gamma^2}{8}k_\parallel^6\right]\right\rangle B_\parallel^2
    -\left\langle a_{22}\left(2\alpha\beta k_\parallel^2+\alpha\gamma k_{\parallel}^4/2\right)\right\rangle B_x B_y.
\end{equation}
Here we took into account that since $a_{22}$ is isotropic, $\langle a_{22} k_x^2\rangle=\langle a_{22} k_y^2\rangle=\langle a_{22} k_\parallel^2\rangle/2$, $\langle a_{22} k_x^2k_y^2\rangle=\langle a_{22} k_\parallel^4\rangle/8$, and $\langle a_{22} k_x^2k_y^4\rangle=\langle a_{22} k_x^4 k_y^2\rangle=\langle a_{22} k_\parallel^6\rangle/16$.

Summing up all the contributions, the conductivity can be written in the phenomenological form,
\begin{equation}
    \sigma = \sigma_{\rm iso}+\delta\sigma_{\rm aniso},
\end{equation}
where
\begin{eqnarray}
    &\sigma_{\rm iso}&=\sigma^0+\left\langle a_{20}\left[(\alpha^2+\beta^2)k_\parallel^2+\frac{\beta\gamma}{2}k_{\parallel}^4+\frac{\gamma^2}{8}k_\parallel^6\right]\right\rangle \\
    &+&\left\langle a_{02}+ \frac{a_{22}}{2}\left[(\alpha^2+\beta^2)k_\parallel^2+\frac{\beta\gamma}{2}k_{\parallel}^4+\frac{\gamma^2}{4}k_\parallel^6\right]\right\rangle B_\parallel^2 \nonumber
\end{eqnarray}
is the isotropic contribution and
\begin{equation}
    \delta\sigma_{\rm aniso} =\left\langle -a_{22}\left(\alpha\beta k_\parallel^2+\frac{\alpha\gamma}{4}k_{\parallel}^4\right)\right\rangle B_\parallel^2 \cos(2\phi)
\end{equation}
is the correction accounting for the in-plane anisotropy. The angle $\phi=\angle B_\parallel$ defines the direction of the in-plane magnetic field with respect to the $[110]$ crystallographic direction.

The angular dependence of $\delta\sigma_{\rm aniso}\propto\cos(2\phi)$ is in qualitative agreement with both the experimental data and numerical simulations, and confirms that the coexistence of Rashba and Dresselhaus SOCs is responsible for the two-fold symmetry of the in-plane anisotropy of the conductivity. Note that the anisotropy vanishes when either $\alpha=0$ (only Dresselhaus SOC is present) or $\beta=\gamma=0$ (only Rashba SOC is present).

From the analysis above, we find the following phenomenological expression for the anisotropy ratio plotted in Fig.2 of the main text,
\begin{equation}\label{eqn:a-ratio}
    \frac{\Delta\sigma(0)-\Delta\sigma(\phi)}{\Delta\sigma(0)}\approx\frac{\left\langle -a_{22}\left(\alpha\beta k_\parallel^2+\frac{\alpha\gamma}{4}k_{\parallel}^4\right)\right\rangle B_\parallel^2}{\sigma_{\rm iso}}[1-\cos(2\phi)].
\end{equation}
This expression properly describes the angular dependence of the anisotropy ratio, as shown in Fig.~2 of the main text.

\section{Electronic Transport}
\begin{figure}[ht]
    \includegraphics[width=0.5\linewidth]{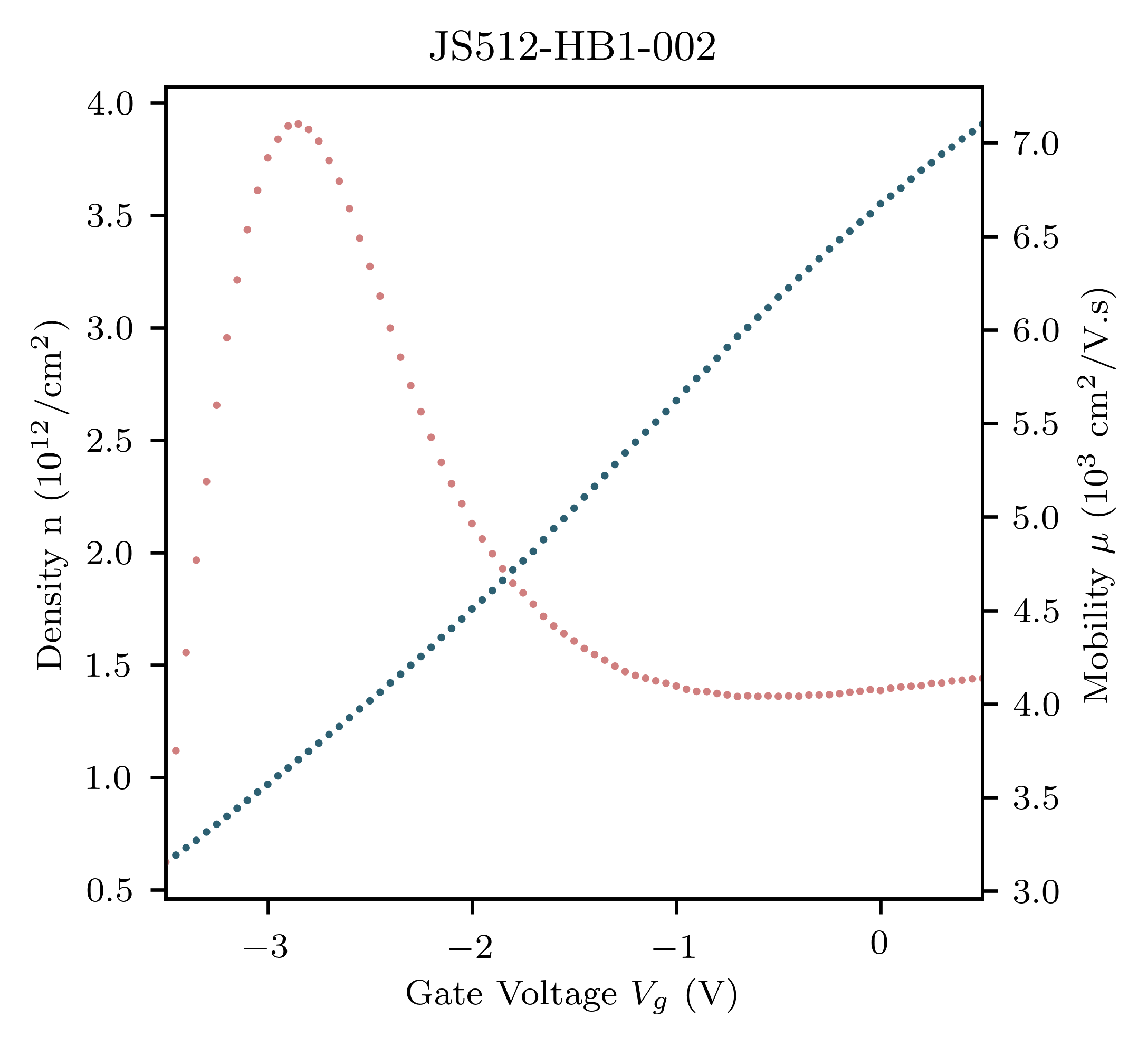} 
    \caption{\label{fig:transport} 
        Transport properties of the two-dimensional electron gas, namely electron density $n$ and mobility $\mu$ as a function of the gate voltage $V_g$. The gate dielectric is a 30 nm layer of Al$_2$O$_3$.}
\end{figure}

Transport properties of the two-dimensional electron gas are obtained from Hall measurements. The electron density is calculated from the Hall resistance as it is proportional to the applied magnetic field $n = B/eR_{xy}$. The mobility is calculated from the longitudinal resistance as $\mu = 1/neR_{xx}G$ where $G$ is the geometric factor of the Hall bars. 
Figure \ref{fig:transport} shows the density and the mobility of the sample as a function of the gate voltage. 
These measurements allow us to calculate other quantities such as the Fermi momentum $k_\mathrm{F} = \sqrt{2\pi n}$, the momentum relaxation time $\tau = m^*\mu / e$ where $m^*$ is the electron's effective mass \cite{2020_Yuan}, and the mean free path $\ell = v_\mathrm{F} \tau $ where $v_\mathrm{F} = \hbar k_\mathrm{F} / m^*$ is the Fermi velocity. 

\section{Orientation Dependence Data and Model}
\begin{figure}[!htb]
\begin{tabular}{cc}
    \centering
    \includegraphics[width=0.45\linewidth]{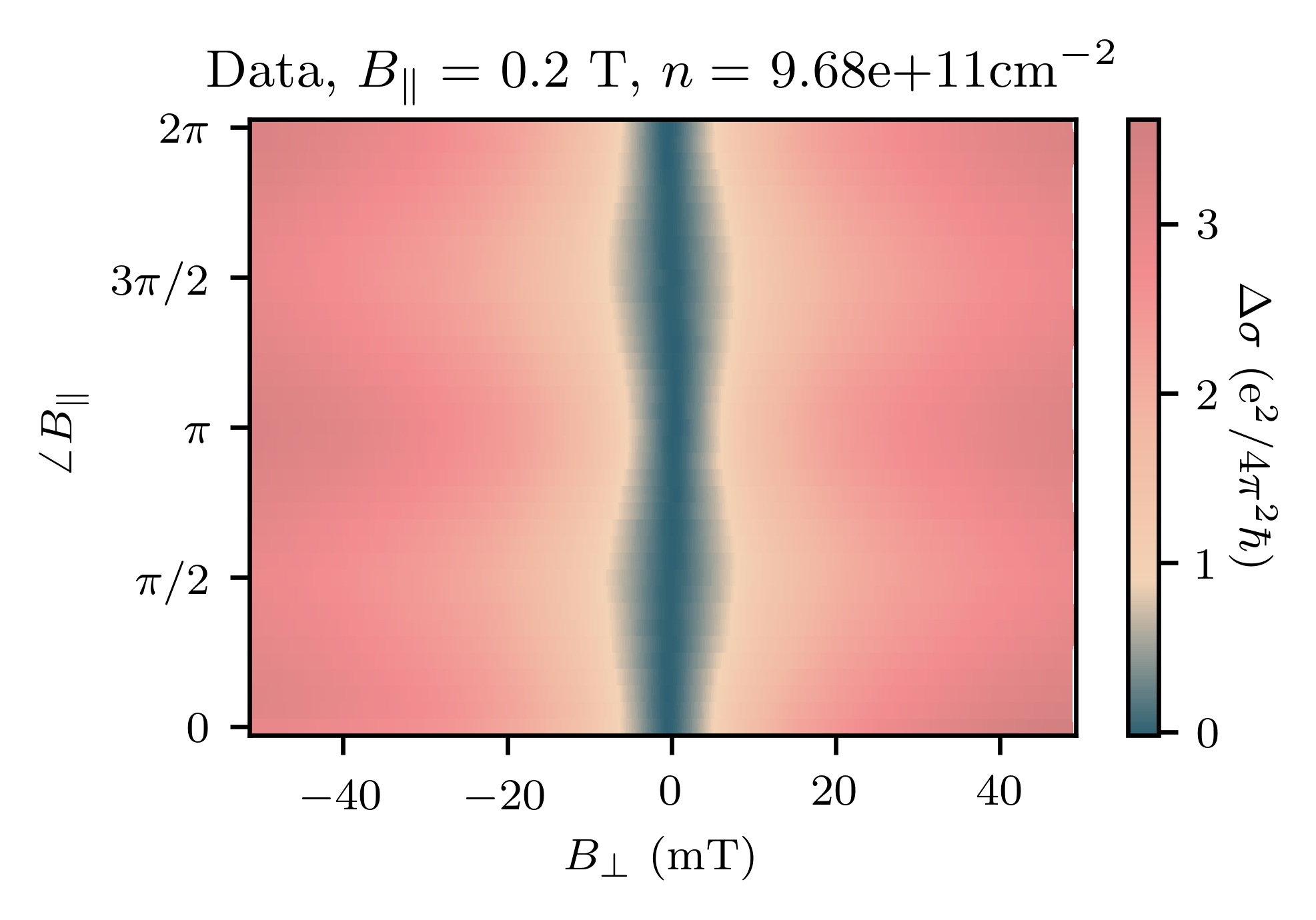} & 
    \includegraphics[width=0.45\linewidth]{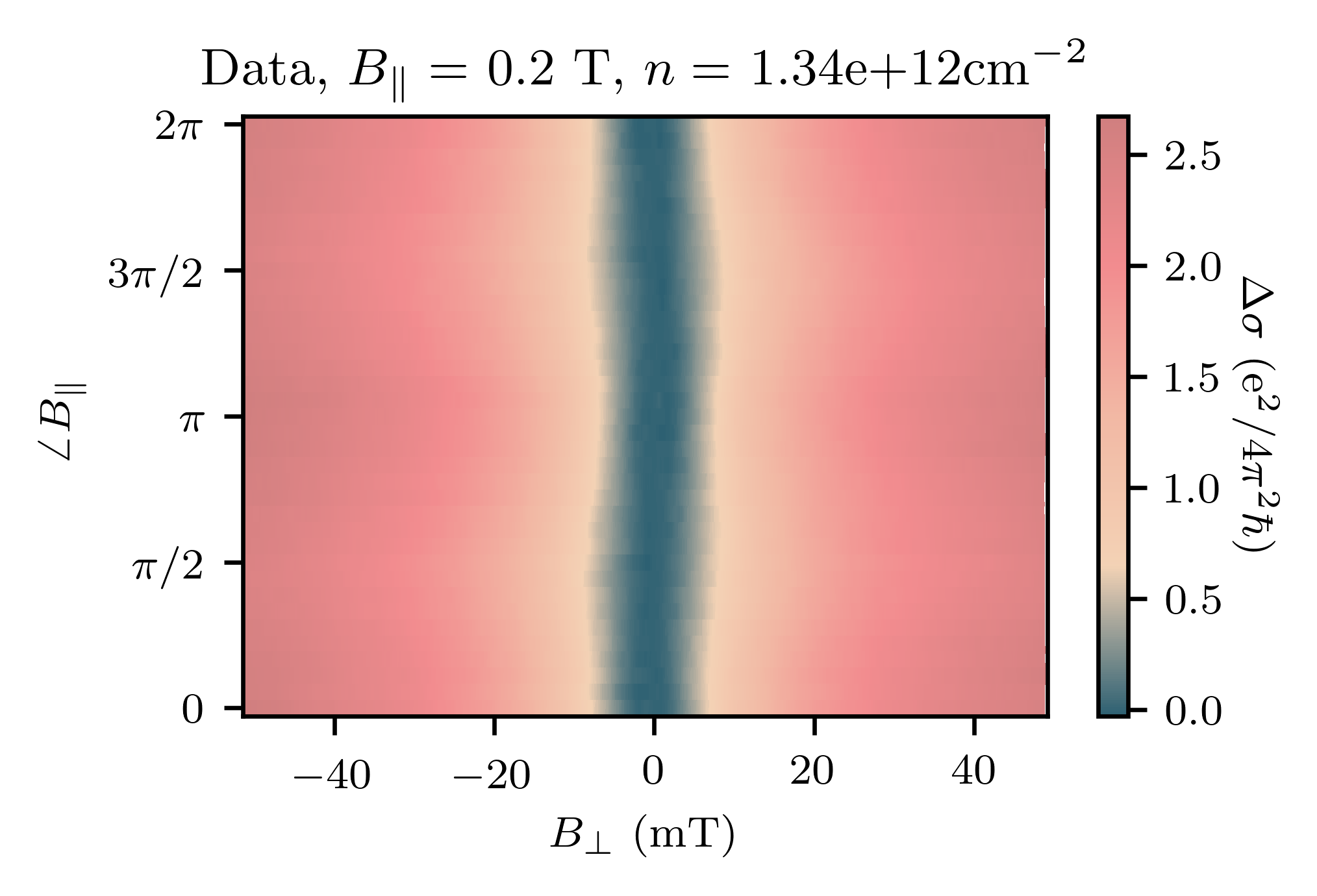} \\
    \includegraphics[width=0.45\linewidth]{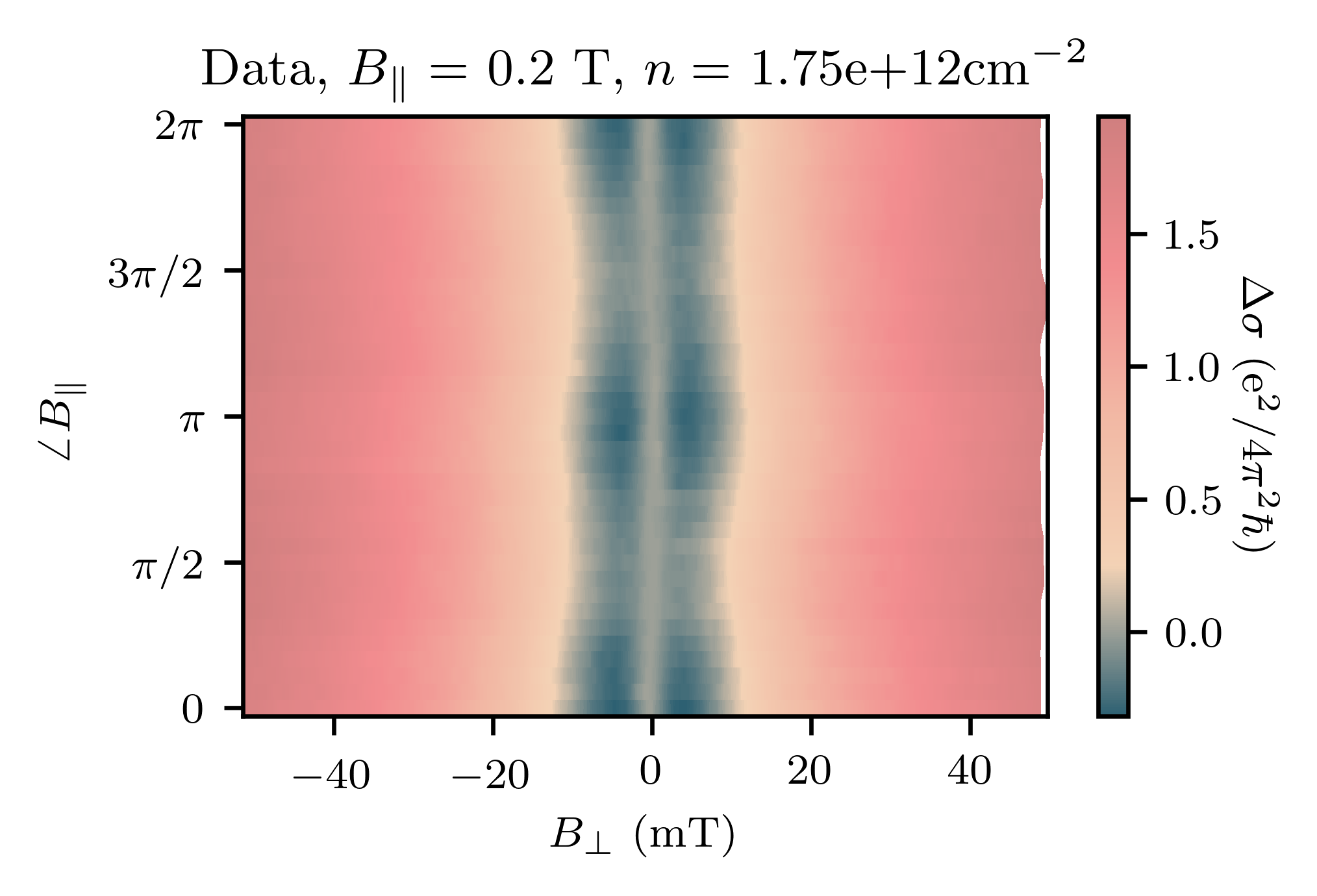} & 
    \includegraphics[width=0.45\linewidth]{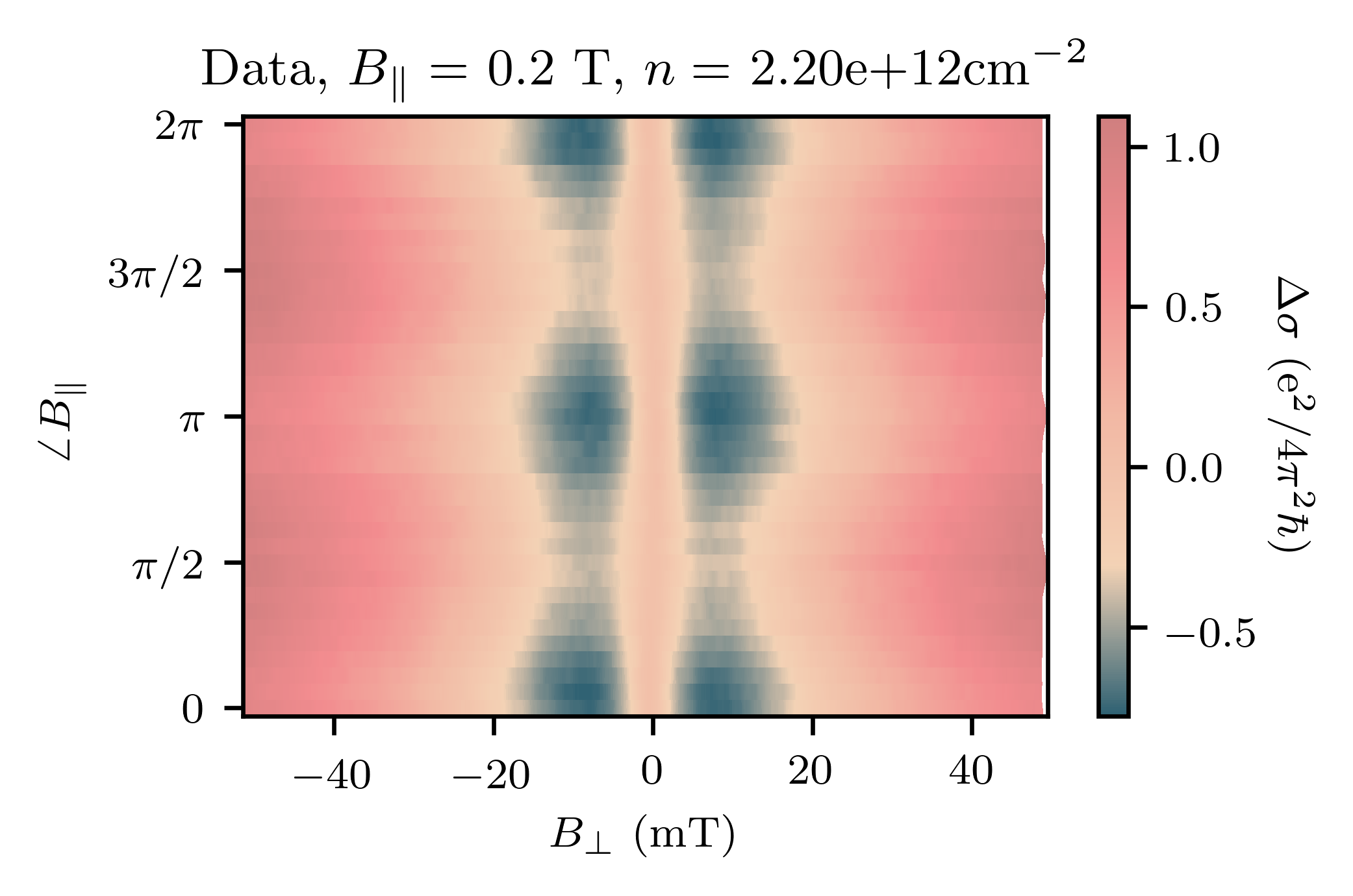} \\
    \includegraphics[width=0.45\linewidth]{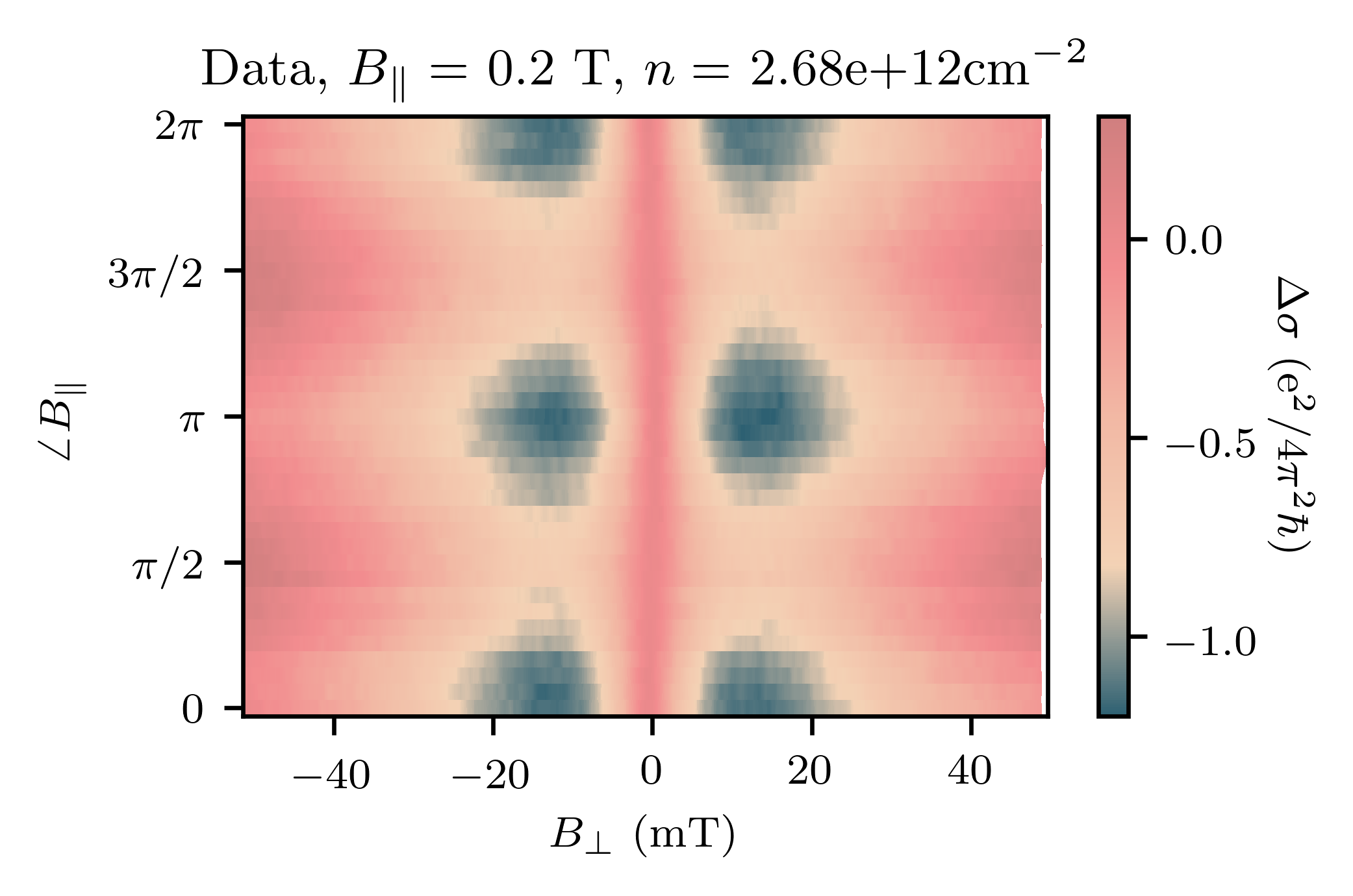} & 
    \includegraphics[width=0.45\linewidth]{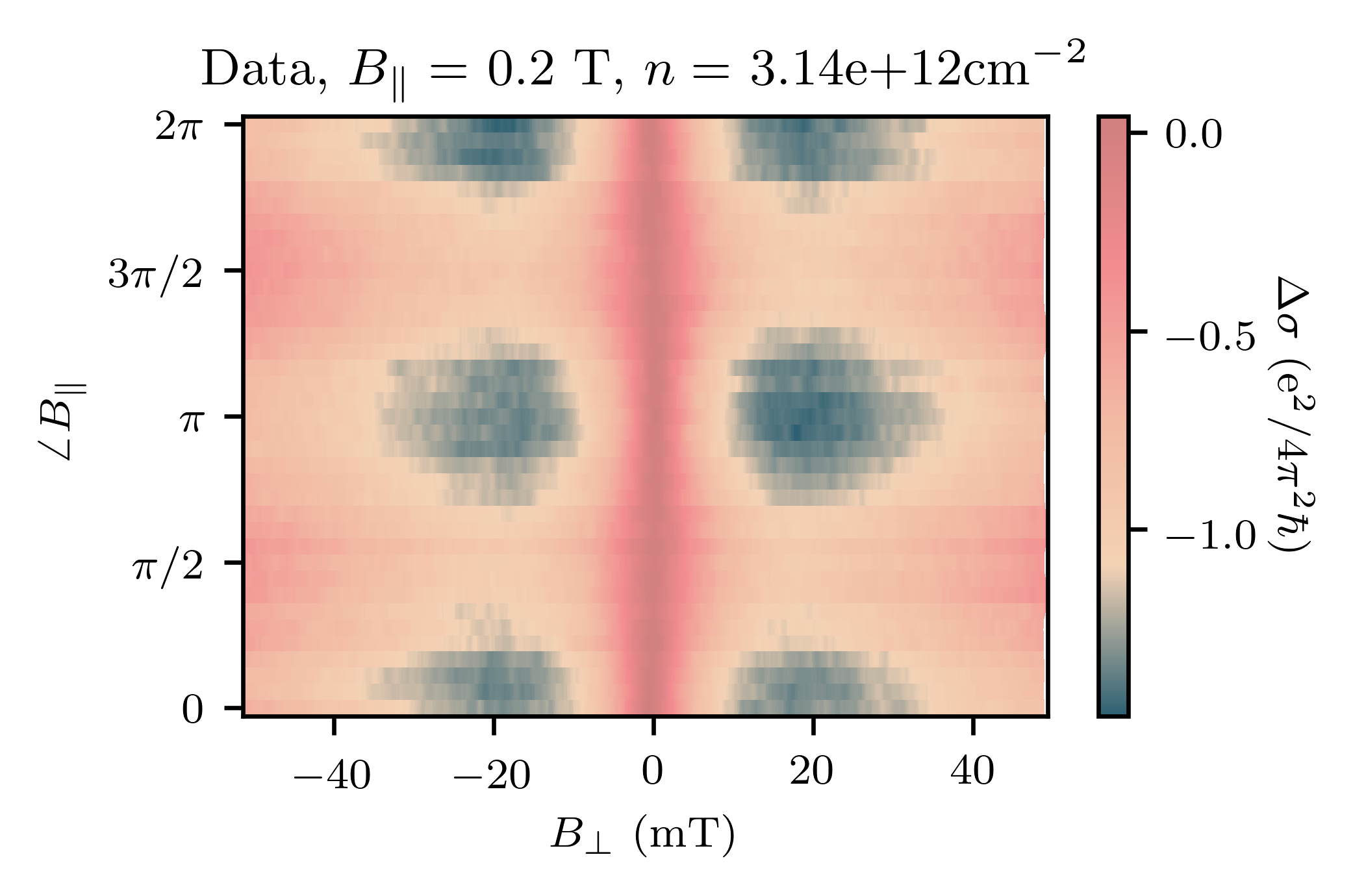} \\
    \includegraphics[width=0.45\linewidth]{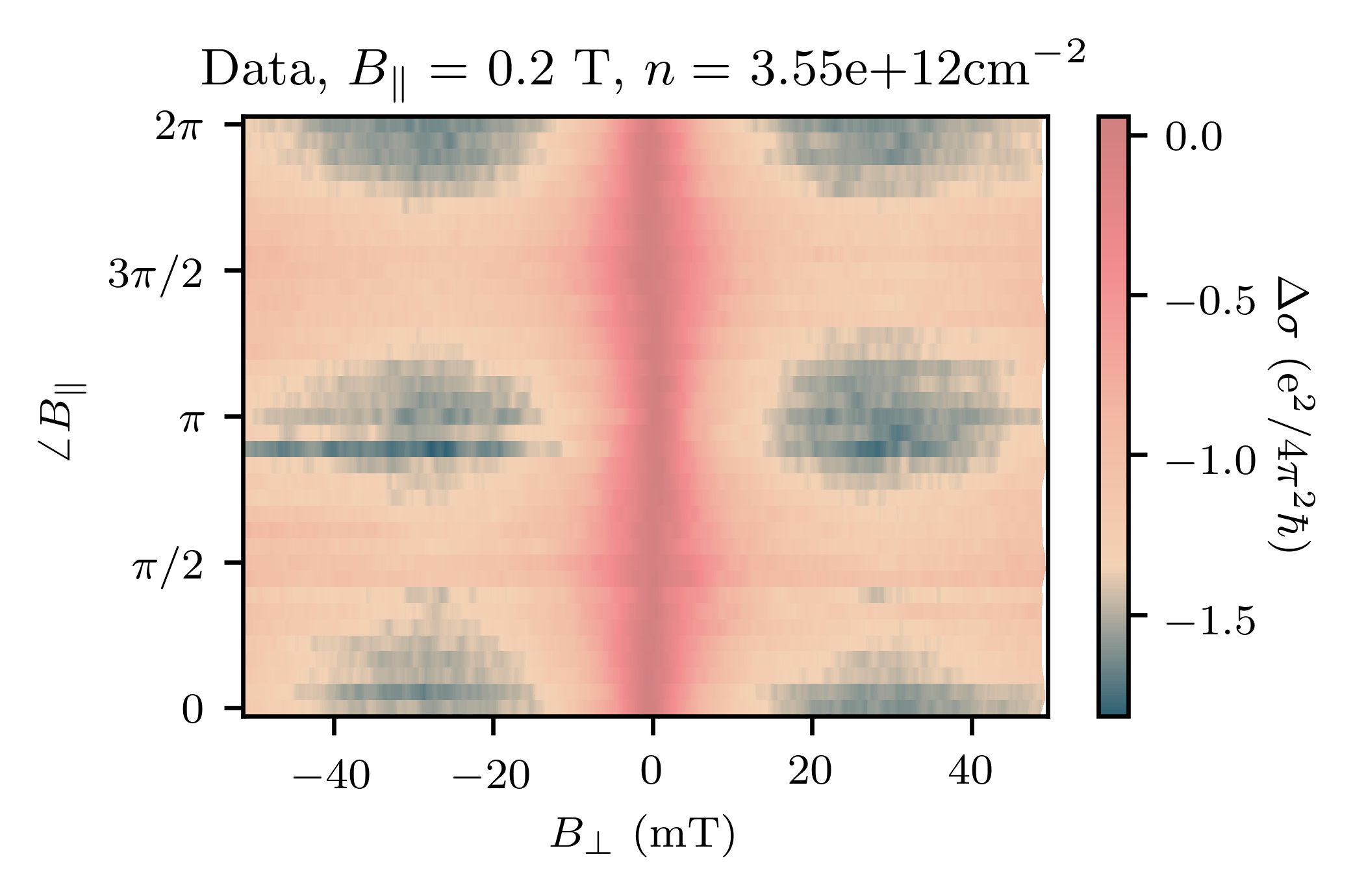} & 
    \includegraphics[width=0.45\linewidth]{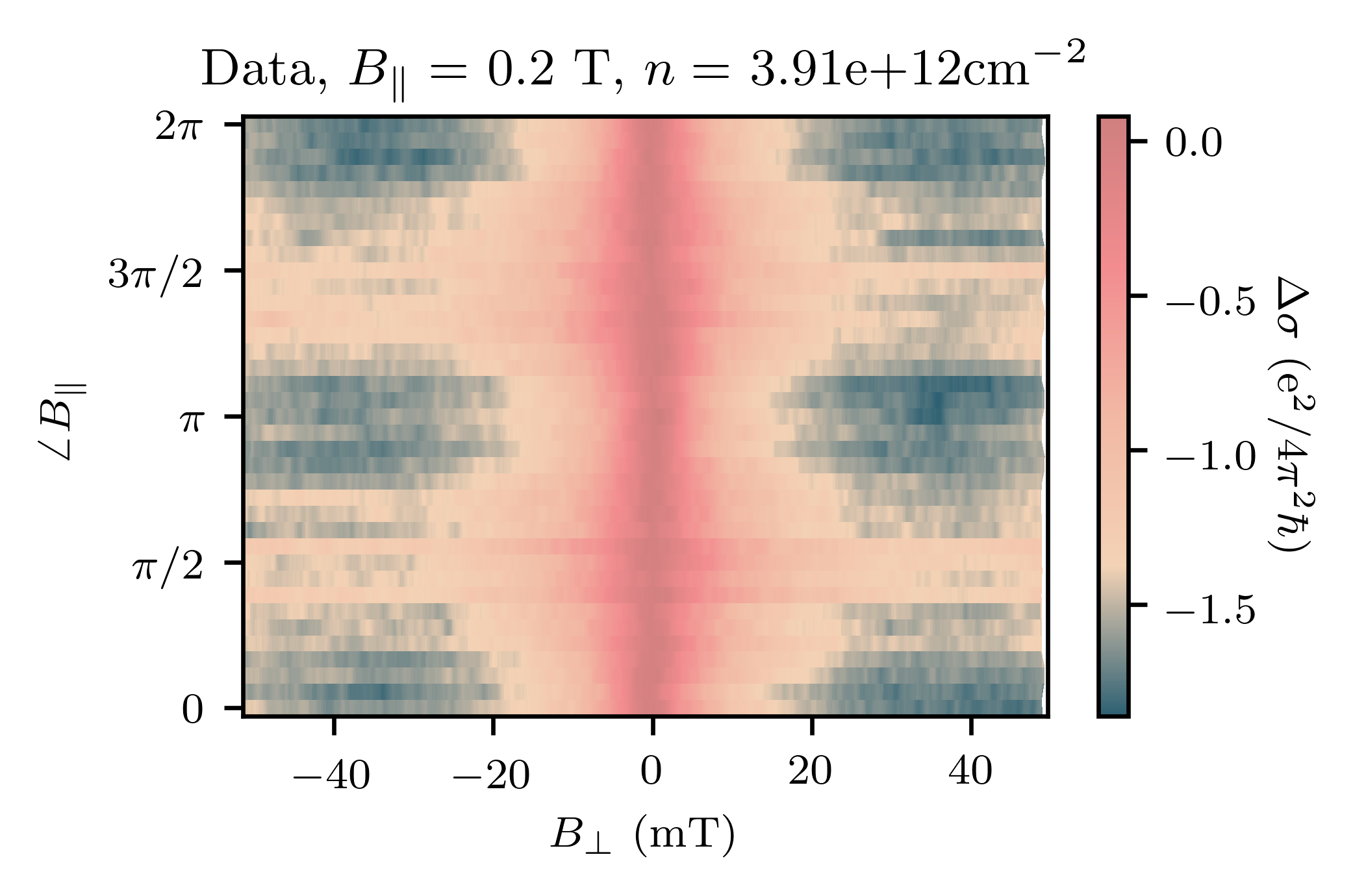} 
\end{tabular}
    \caption{\label{fig:supp_angle} Angle dependence data. Magnetoconductivity versus the angle of the in-plane magnetic field with a fixed magnitude of $B_\parallel = 0.2$ T. Different plots correspond to different densities (gate voltages). The angle $\angle B_\parallel$ is with respect to the [110] crystal axis. A horizontal cut on each plot gives the weak antilocalization signature. The data is manually shifted so that $\Delta \sigma = 0$ at $B_\perp = 0$.}
\end{figure}

\begin{figure}[!htb]
\begin{tabular}{cc}
    \centering
    \includegraphics[width=0.45\linewidth]{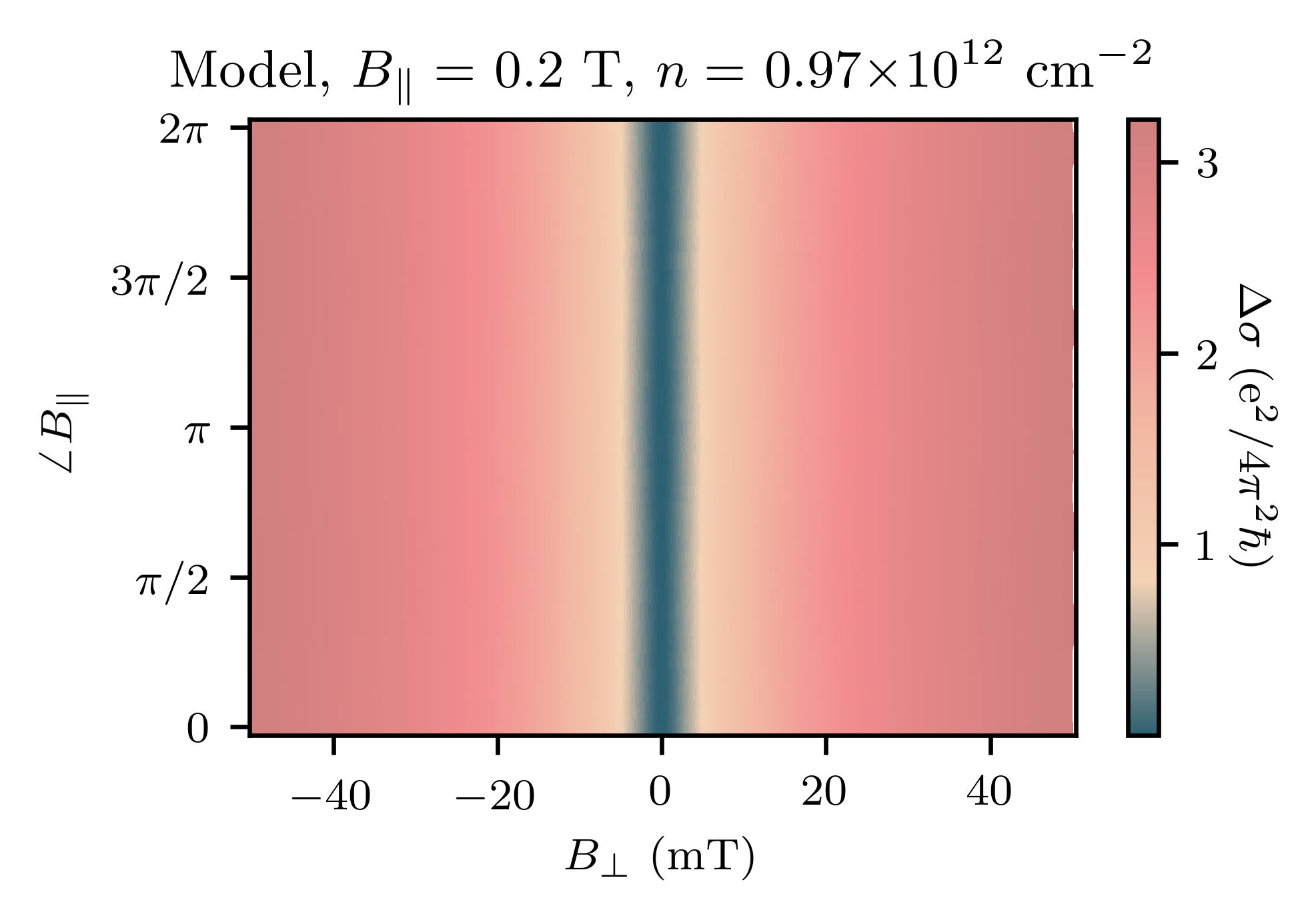} & 
    \includegraphics[width=0.45\linewidth]{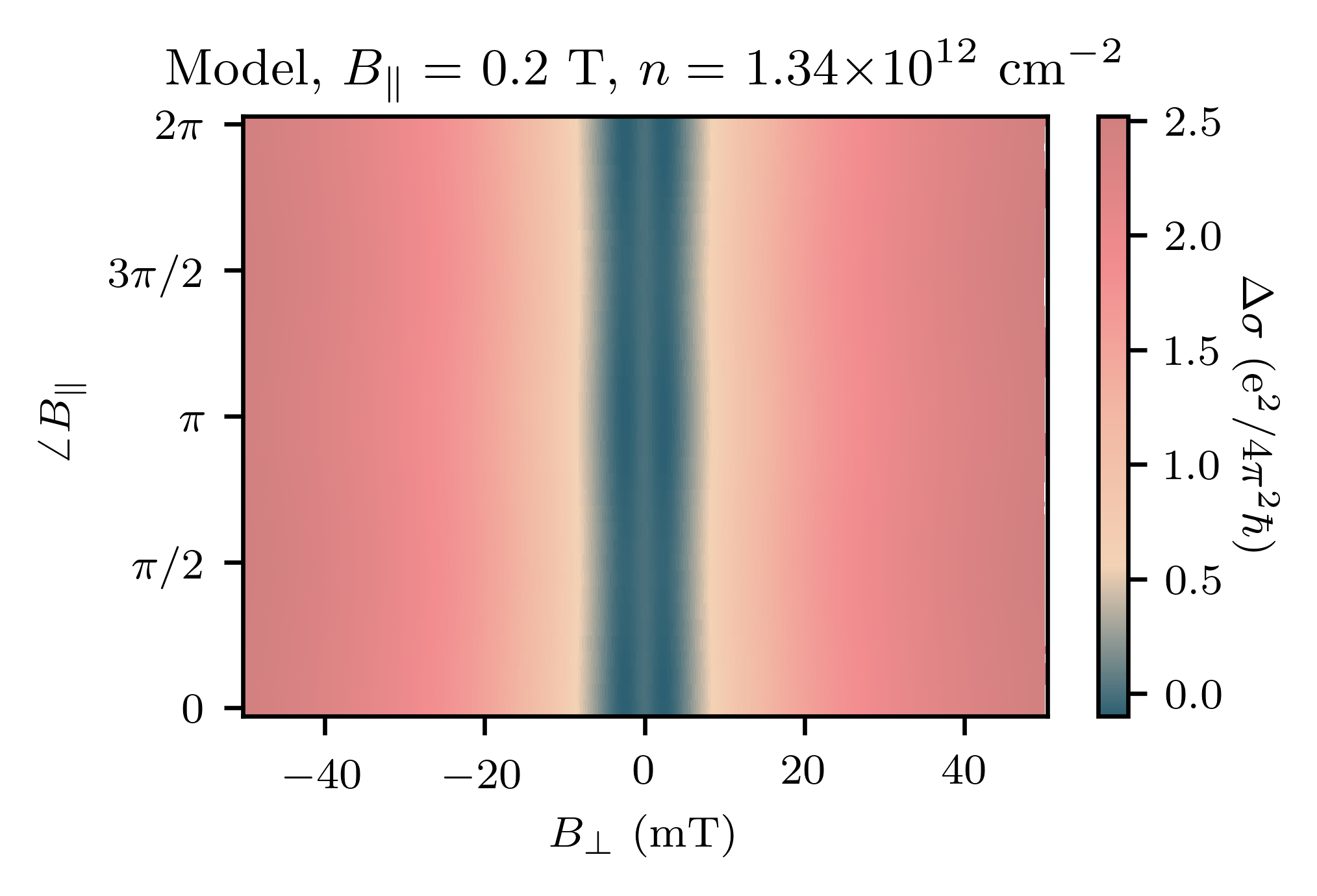} \\
    \includegraphics[width=0.45\linewidth]{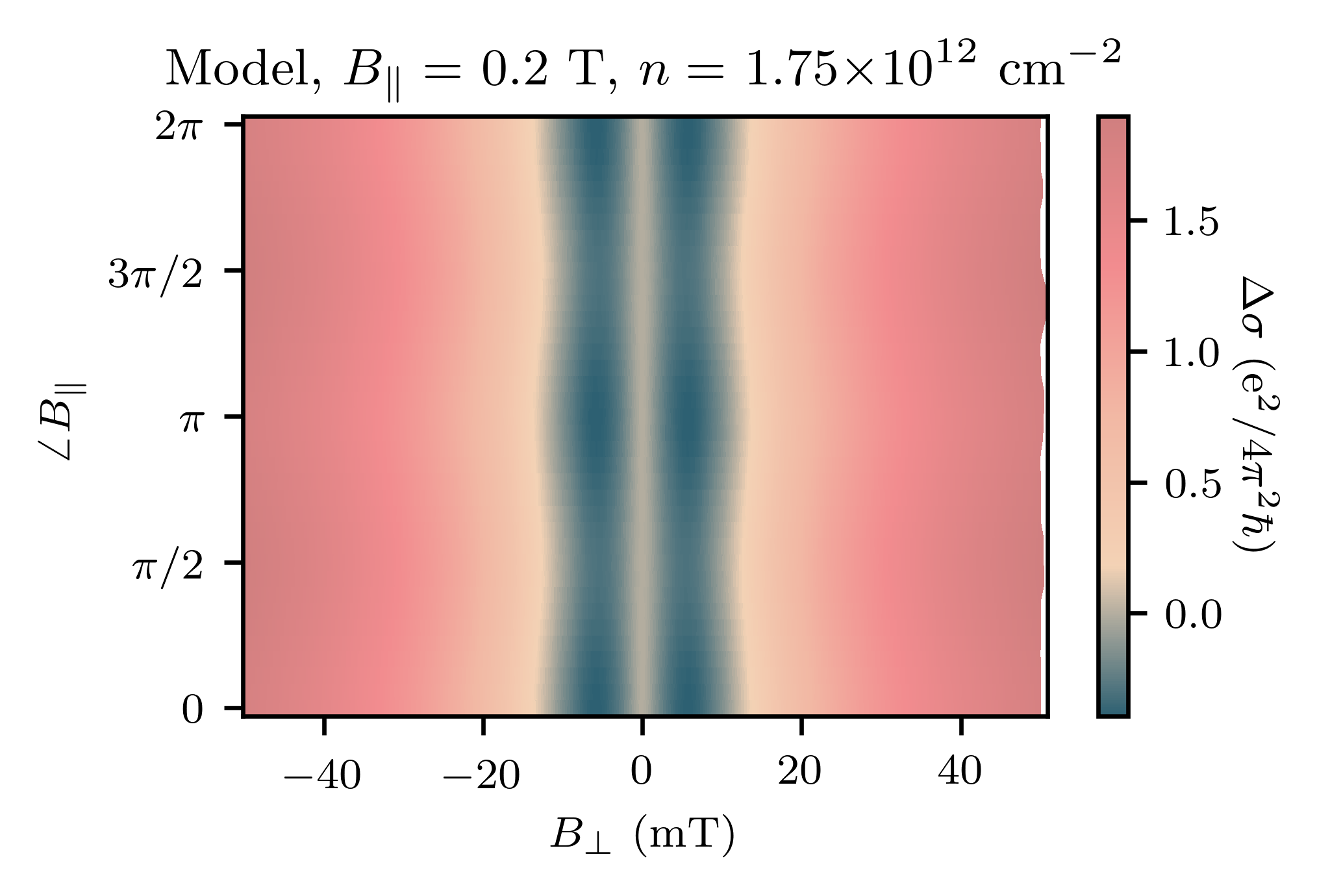} & 
    \includegraphics[width=0.45\linewidth]{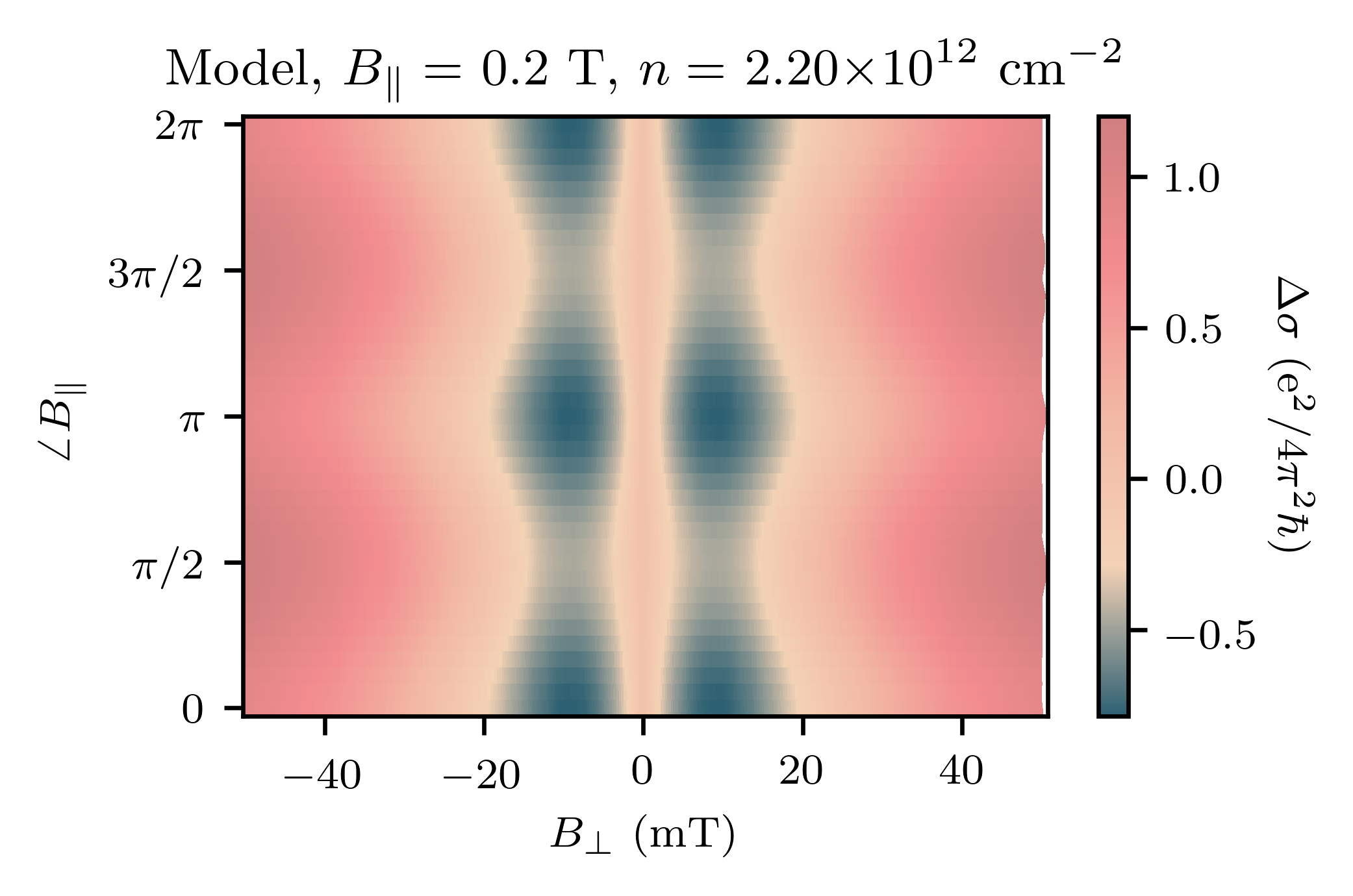} \\
    \includegraphics[width=0.45\linewidth]{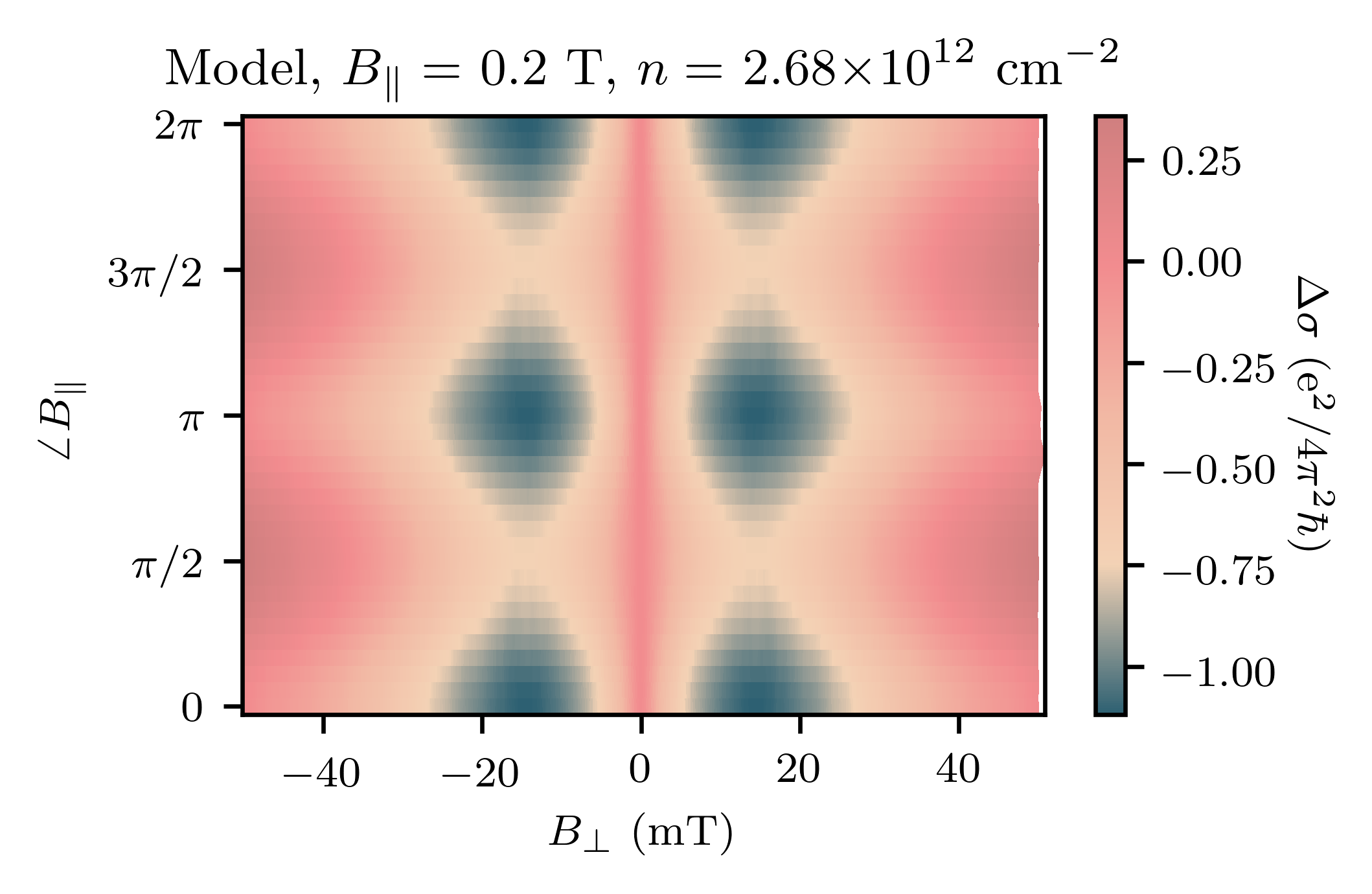} & 
    \includegraphics[width=0.45\linewidth]{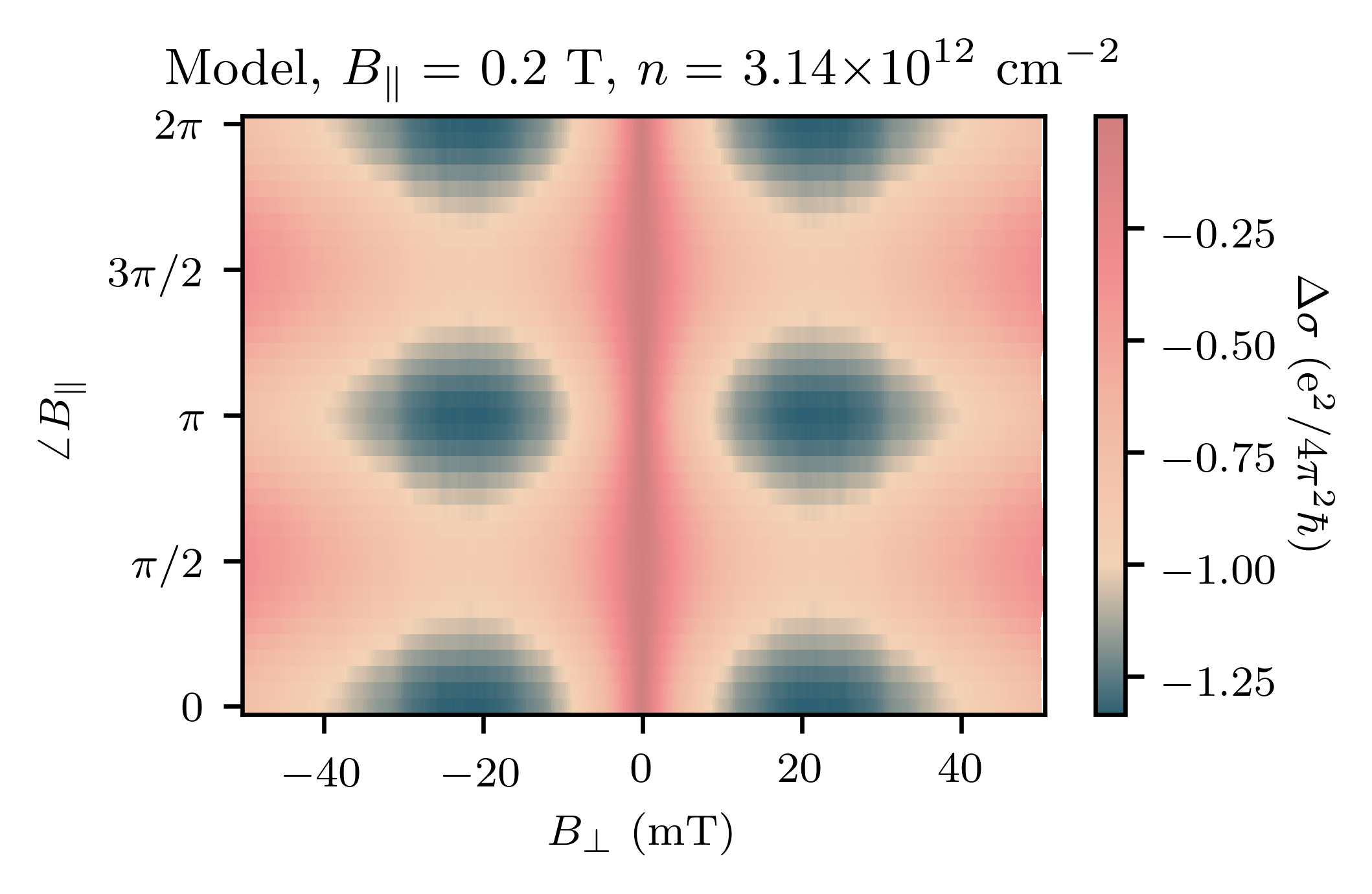} \\
    \includegraphics[width=0.45\linewidth]{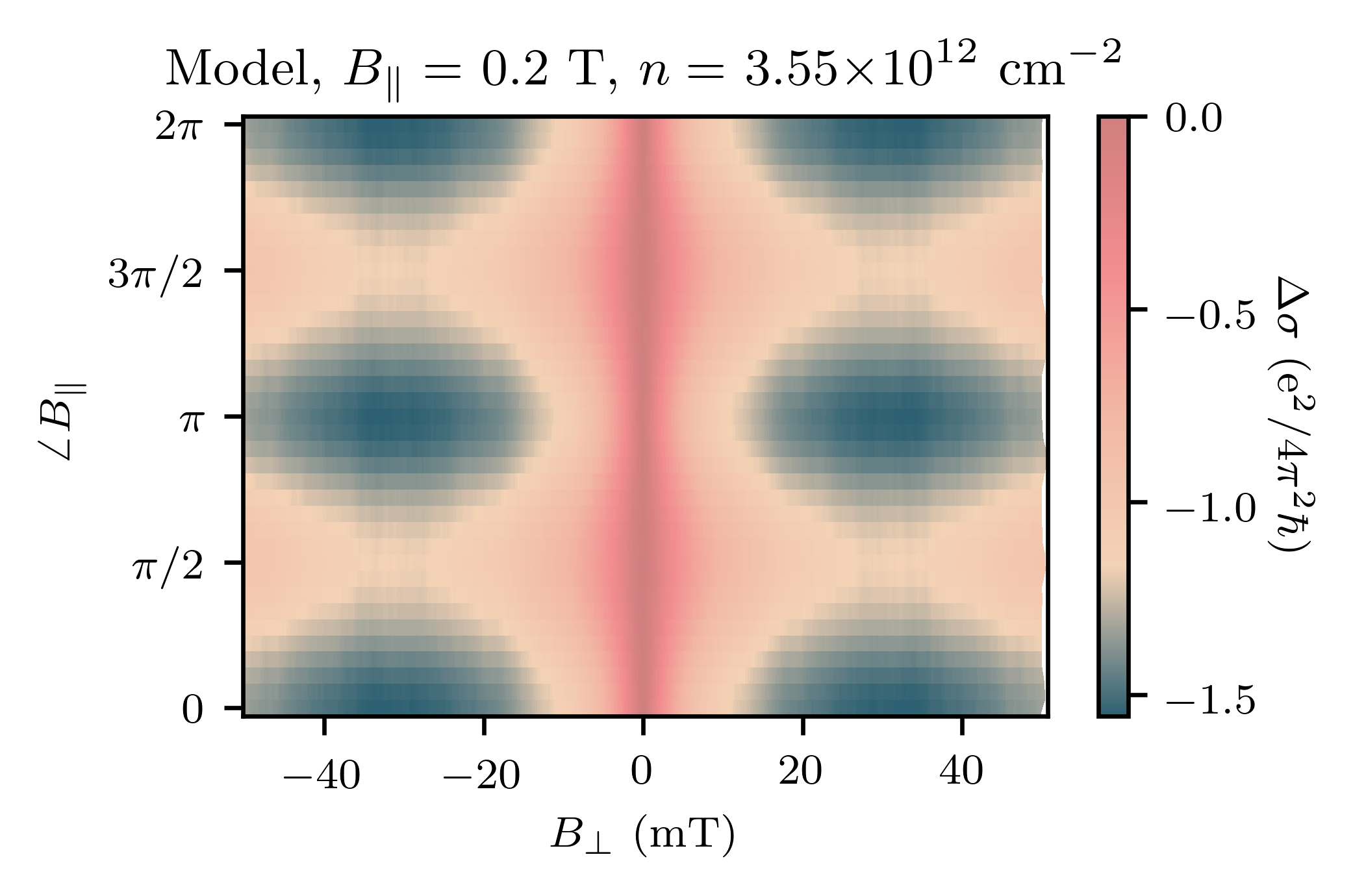} & 
    \includegraphics[width=0.45\linewidth]{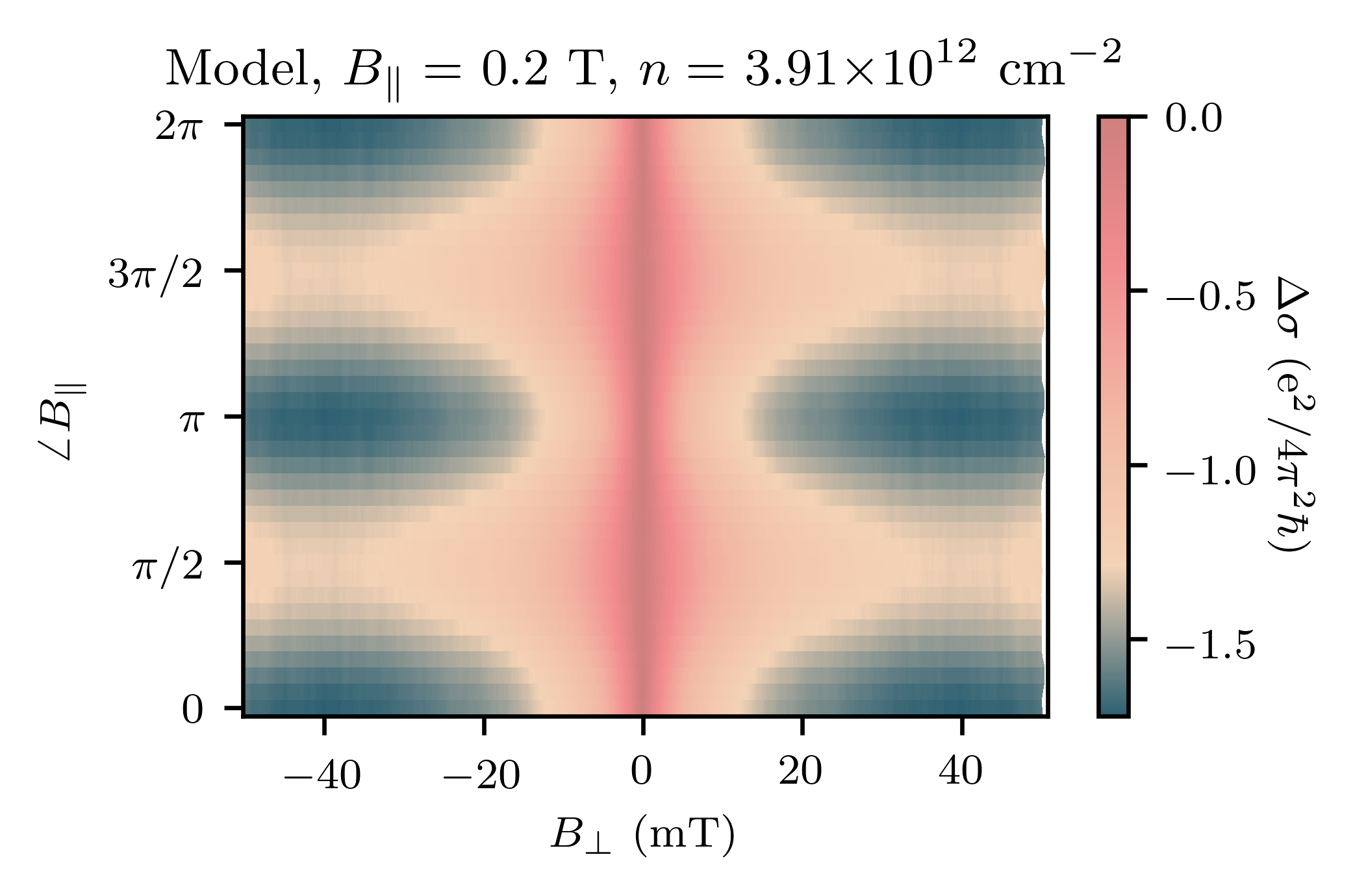} 
\end{tabular}
    \caption{\label{fig:supp_angle_sim} Magnetoconductivity reproduced using the semiclassical theory of weak antilocalization at densities corresponding to Fig. \ref{fig:supp_angle}}
\end{figure}

The complete set of the orientation dependence measurements, that is the magnetoconductivity $\Delta \sigma$ vs the orientation of the in-plane field $\angle B_\parallel$, is provided Figure \ref{fig:supp_angle}. 
Different plots correspond to different densities. 
The origin of the anisotropy is assumed to be due to the combination of Rashba and Dresselhaus spin-orbit coupling which reduces the rotational symmetry in the plane of the 2D system to a two-fold symmetry.
Figure \ref{fig:supp_angle_sim} illustrates the corresponding magnetoconductivity generated numerically from the semiclassical weak antilocalization model. 
The model captures the anisotropy since both Rashba and Dresselhaus terms are non zero. 
Without introducing the Dresselhaus term, one needs to resort to an anisotropic Zeeman splitting to explain the experimental data. This is not desirable since, as we showed in Section \ref{sec:aniso}, an anisotropic Zeeman splitting can only be explained using higher order perturbation terms which are estimated to be negligible. 

\end{document}